\documentclass{article}

\usepackage{arxiv}

\usepackage[utf8]{inputenc} 
\usepackage[T1]{fontenc}    
\usepackage{hyperref}       
\usepackage{url}            
\usepackage{booktabs}       
\usepackage{amsfonts}       
\usepackage{nicefrac}       
\usepackage{microtype}      
\usepackage{graphicx}
\usepackage{natbib}
\usepackage{doi}

\title{Activity-Based Recommendations for Demand Response in 
	Smart Sustainable Buildings}

\author{\href{https://orcid.org/0000-0003-3506-4744}{\includegraphics[scale=0.06]{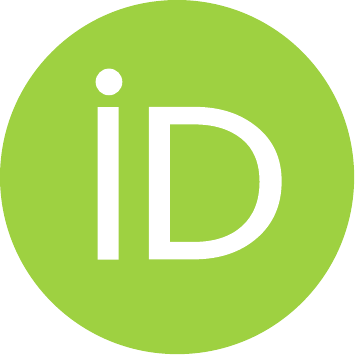}\hspace{1mm}Alona Zharova\thanks{Correspondence author.} } \\
	Humboldt-Universität zu Berlin\\
	Berlin, Germany \\
	\texttt{alona.zharova@hu-berlin.de} \\
	\And
	Laura Löschmann \\
	Humboldt-Universität zu Berlin\\
	Berlin, Germany \\
	\texttt{laura.loeschmann@hu-berlin.de} \\
}

\renewcommand{\shorttitle}{Activity Prediction for a Recommendation System}

\hypersetup{
pdftitle={titel},
pdfsubject={subject},
pdfauthor={authors},
pdfkeywords={First keyword, Second keyword, More},
}

\begin{document}
\maketitle

\begin{abstract}
	The energy consumption of private households amounts to approximately 30\% of the total global energy consumption, causing a large share of the CO2 emissions through energy production. An intelligent demand response via load shifting increases the energy efficiency of residential buildings by nudging residents to change their energy consumption behavior. This paper introduces an activity prediction-based framework for the utility-based context-aware multi-agent recommendation system that generates an activity shifting schedule for a 24-hour time horizon to either focus on CO2 emissions or energy costs savings. 
	In particular, we design and implement an Activity Agent that uses hourly energy consumption data. It does not require further sensorial data or activity labels which reduces implementation costs and the need for extensive user input. Moreover, the system enhances the utility option of saving energy costs by saving CO2 emissions and provides the possibility to  focus on both dimensions.
	The empirical results show that while setting the focus on CO2 emissions savings, the system provides an average of 12\% of emissions savings and 7\% of cost savings. When setting the focus on energy costs savings, 20\% of energy costs and 6\% of emissions savings are possible for the studied households in case of accepting of all recommendations.
	Recommending an activity schedule, the system uses the same terms residents describe their domestic life with. Therefore, recommendations can be more easily integrated into daily life supporting the acceptance of the system in a long-term perspective. 
\end{abstract}

\keywords{Activity Prediction \and Recommendation System \and Demand Response \and Energy Efficiency \and CO2 Emissions}

\section{Introduction}

Climate change confronts the humanity with complex global issues involving many dimensions [1]. This makes it one of the greatest challenges, requiring the interaction of the most diverse areas under the same goal. The European Union aims at becoming climate neutral by 2050 [2] that is either achieved by activities that must release no greenhouse gases at all [3] or that all human-made greenhouse gas emissions have to be removed from the atmosphere and permanently sequestered through, for instance, reforestation or carbon capture technologies [4]. However reliable and sustainable technologies that are cost-efficient and able to sequester CO2 on a large scale have not yet been realized [3]. 

The energy sector is responsible for 60\% of the global man-made greenhouse gas emissions which makes it the main factor of climate change [5]. Following industry and transport, private households are the third largest energy consumers, accounting for approximately 30\% of the total energy consumption and with this, are responsible for a large share of CO2 emissions [6]. The increase of the share of renewable energy within the global energy mix is inevitable in the fight against climate change. However, the production of green energy is dependent on weather conditions. This can result in an imbalance between the supply of renewable energies and the demand for energy, thus, causing difficulties for the grid’s stability. Ideally, green energy should be either consumed when it is available or stored for future consumption. 

Until now, extending energy storage is relatively expensive and there is still ongoing research on efficient technologies storing energy over longer durations [7]. Therefore, increasing the efficiency of energy consumption through intelligently leveraging the consumption, for instance through load shifting, is an effective, quick to implement and affordable way to stabilise the energy system and reduce CO2 emissions. For the practical implementation of load shifting, private households require data-driven decision support. Recommendation systems provide the needed framework by suggesting energy-efficient actions using the household’s individual data.

Previous works explore recommendation systems for demand response of various architectures applying several designs based on, for instance, semantic modelling [9], fuzzy logic [10] or multi-agent systems [11], [12], [13] and using various types of data such as energy consumption data, price data, energy production data or data measured by temperature, light or motion sensors. 
So far, recommendation systems developed for load shifting purposes mainly focussed on recommending the shift of the usage of a device to save energy costs [9], [10], [11],[12], [13]. There is only limited work in activity-aware systems and recommending the shifting of domestic activities corresponding to energy consuming devices [14, 15]. The works approaching this problem, however, use different sensor data with activity labels. Due to the high amount of sensors needed, this increases the implementation effort and costs in real-life applications  and burdens users with the requirement of tracking their activities for a sufficient amount of time.

This paper introduces a human activity-driven approach to generating recommendations for load shifting. The recommendation system provides the option to focus the activity shifting on CO2 emissions or energy costs savings or both. As a utility-based system, it models user preferences depending on user availability, device usage habits and the household’s individual activities and considers the user’s motivation regarding saving CO2 emissions or energy costs. As a context-aware system, it requires external hourly CO2 emissions and hourly price signals for the following day as well as the household’s energy consuming activities, the devices used to carry out the activities and historical energy consumption data on device level. Using this data, the recommendation system generates an activity schedule containing the best activity starting hours along with the durations of the activities that are predicted for the next 24 hours with an hourly precision that would result in the highest possible savings taking into account the availability and devise usage habits of the user as well as the user’s focus for savings. The system is built using the multi-agent architecture which provides flexibility and be can easily adjusted and enhanced.

Recommending an activity schedule, the system uses the same terms residents describe their domestic life with. Therefore, recommendations can be easier integrated into daily life which supports the acceptance of the system over a long period. The Activity Agent provides activity forecasts on an hourly basis and with this enables multiple predictions of the same activity within one recommendation period. This also facilitates a more precise calculation of costs per activity since individual duration for each predicted activity can be forecasted. Recommending an activity schedule reduces the number of recommendations per recommended time horizon compared to recommending a device usage schedule since activities represent the usage of multiple devices whose recommended shifting is combined through the activity. We suggest an Activity Agent that uses hourly energy consumption data. It does not require further sensorial data or activity labels which reduces implementation costs and the need for extensive user input. Moreover, the system enhances the utility option of saving energy costs by saving CO2 emissions and provides the possibility to  focus on both dimensions. The consumer can also change the focus of the system while using it. 

The empirical analysis aims at quantifying how much CO2 emissions and energy costs a user can save by applying the proposed system. To this end, recommendations are created over a period of one year to calculate the emissions and costs for executing the activities while accepting and rejecting the recommendations to receive the possible savings. 
The results show that while setting the focus on CO2 emissions savings, the system provides an average of 12\% of emissions savings and 7\% of cost savings. When setting the focus on energy costs savings, 20\% of energy costs and 6\% of emissions savings are possible for the studied households in case of accepting of all recommendations.

The contributions of this paper can be summarized as follows. First, we design and implement an Activity Prediction Agent as part of a multi-agent recommendation system using activities for load shifting in private households. Second, the system solves a complex task of predicting an activity schedule while requiring a minimal user input and unlabeled data of only one sensor type. Third, the system provides the flexibility of choosing the utility among saving CO2 emissions and energy costs.

The remainder of this paper is structured as follows. Section 2 provides an overview of related literature on recommendation systems for demand response as well as activity recognition and prediction in the context of energy consumption. Section 3 describes the methodology of the proposed system and the framework for its performance evaluation. In section 4, we test the system using real-world data to analyse and discuss its performance. Section 5 elaborates the implications, contributions, limitations and future work, while section 6 concludes the paper.

\section{Literature Review}

\subsection{Demand Response}

Demand response aims at an efficient energy usage on the demand side to better adjust the power demand with the supply. It comprises approaches to encourage changes in electric usage by end-use customers from their normal consumption patterns in response to changes of electricity prices or CO2 emissions through energy production over time or periods in which an incentive is considered for voluntary demand reduction [16]. These methods provide customers with a significant role in the operation of the electric grid. Demand response programs can lower the cost of electricity in wholesale markets which in turn lead to lower retail prices or are able to lower CO2 emissions during electricity production [17].
Demand response programs are entirely dependent on the participation of the consumers [18]. Implementing a demand response program, the residents’ comfort is crucial to prevent users’ disaffection and the abandonment of the system [19]. Therefore, the programs have to be convenient for the customer providing them with benefits at short notice. These benefits can be in the form of a decrease of their energy bill due to time of use pricing or other financial rewards, or a sufficient reduction of their ecological footprint.


Demand response programs in private households can effectively be implemented via an intelligent home energy management system (HEMS) that comprises sensing and measuring devices (smart meter, sensors for motion, temperature or light), the central platform analyzing and optimizing the energy consumption as well as a user interface to enable interaction between the residents and the HEMS. Energy consumption data on appliance level is necessary for the control algorithm to manage the energy usage. Smart appliances, low-cost smart plugs or load disaggregation techniques such as Non-Intrusive Load Monitoring (NILM) provide such data [22]. 

\subsection{Recommendation Systems in Smart Homes}

To actively participate in demand response programs, consumers need information regarding the times when energy prices are low or their specific incentives are met. Recommendation systems can support energy consumers in their private homes to adapt their energy usage. 
Recommendation systems are algorithms that suggest relevant items. Due to the virtually infinite variety of choices online, recommendation systems are an essential part in the fields of online advertisement, social networks or e-commerce. In general, there are different types of recommendation systems. The content-based recommendation system filters for similar items based on shared features and characteristics. A recommendation system based on collaborative filtering declares two items as similar if a great number of users that are similar to the target user, buy or rate these two items. Users are similar if they share a great number of preferences [23]. 

Using these recommendation methods to predict a useful activity schedule according to the user’s preferences in a private household for load shifting is not feasible. The recommended item (activity schedule) should have a sufficient complexity, simultaneously taking the user’s preferences into account to be useful which makes content-based recommendation systems unsuitable. Since users' preferences, their activities and used devices are totally individual, collaborative filtering is also not a proper choice for the recommendation system, especially due to privacy concerns since energy consumption data and data about the individual activities would need to be shared between households [13].

In the context of Internet of Things, there are two more types of recommendation systems. A knowledge-based system includes information about the user into its recommendations. This information represents the requirements an item has to meet in order to be recommended. If a user, e.g., is willing to accept greater changes in order to save energy, a recommendation system can provide different recommendations as if the user is hesitating to change their routine. A utility-based system gives information based on the utility for the user [12]. In this context the utility for the user could be to save money which would be the incentive to make bigger changes regarding their energy consumption. A requirement for this form of recommendation system is the awareness of context the user’s utility is focussed on. For instance, to provide recommendations that help the user to reduce the electricity bill, the recommendation system needs to have the contextual information of the electricity price.

Recommendation systems for demand response in smart homes have the goal to combine an efficient energy management by shifting or minimizing power consumption and preventing waste while preserving the comfort of the residents [24]. Marinakis et al. [9] propose a system for intelligent energy management in buildings with semantic modelling integrating all the entities that constitute the environment of a smart building. Using various types of real-time data (e.g., building’s data, energy prices, weather data) and predicted data produced by prediction models (e.g., energy consumption, renewable energy production) the tool gives practical action plans for the building’s residents including tailor-made recommendations and energy saving tips. 
An energy management controller based on fuzzy logic and heuristic optimisation techniques is introduced by Khalid et al. [10] aiming at the reduction of the peak-to-average ratio, the minimization of the total energy consumption and, as a consequence thereof, the reduction of energy costs. Sensors for outdoor temperature, user’s availability and light, energy consumption data per appliance as well as information about price and the demand of energy are utilised to reduce the energy consumption of illumination and HVAC systems as well as provide and schedule for the shiftable load.

A real-time electricity scheduling for home energy management is presented by Li et al. [25] focusing on reducing the energy costs improving the usage of renewable energies assuming the presence of renewable energy systems (RES), such as photovoltaic arrays or wind turbines as well as an energy storage system (ESS). The system deploys a 24-hour rolling horizon and gets predicted information as input which are updated at each prediction period. Therefore, the system focusses on the accuracy for the upcoming part of the rolling horizon, taking a loss of accuracy for the remaining ones. According to the prediction of renewable energies, the utility’s energy prices as well as user’s preferences, shifting schedules for deferrable appliances are calculated.

A recommendation system adopting the architecture of a multi-agent system contributes additional flexibility to the system. A multi-agent system comprises a collection of several intelligent agents, having the possibility to communicate and coordinate their actions, each one aiming to accomplish the overall goal of the recommendation system [26].
Fioretto et al. [11] introduce a smart home device scheduling system that coordinates smart device schedules across a coalition of multiple smart homes as a multi-agent system with each agent controlling a set of appliances and sensors of one smart home aiming at the minimization of energy usage in peak hours.
Jiménez-Bravo et al. [12] implement a multi-agent recommendation system using energy consumption data from smart plugs. The goal of this system is to give the user recommendations in a way that the energy consumption is distributed along different hours of the day as well as to reduce the energy consumption giving the user the incentive to save money. The recommendation system consists of device agents, one agent for one appliance, a crawler module to crawl the relevant energy prices, a recommendation module giving the recommendations by using the information from the device agents and the crawler module and a control agent initializing the prior mentioned agents and modules. 
A utility-based context aware multi-agent recommendation system by Riabchuk et al. [13] generates device-usage schedules for a 24-hour time window based on energy consumption data on appliance level as well as device-specific load profiles taking into account day-ahead energy prices and user’s availability to enable energy cost savings. 

\subsection{Domestic Energy Consumption Activities}

Load shifting by recommending residents when to use which appliance for a most efficient energy use in their private household can be too technical as it does not correspond to the way residents would describe how they spend their time at home. The domestic use of energy is a result of deeply embedded social practices [27]. These social practices can simply be depicted as activities like \textit{Cooking, Cleaning, Working or Laundering} and feature a greater alignment with the residents’ own experiences of domestic life. Providing energy usage schedules based on activities comes across more clearly for the household’s residents to understand and follow them [28]. Activities can comprise the usage of various devices and can therefore slim down the schedule instead of having a usage schedule for every appliance in the household. Furthermore, activities form a more stable description of life at home, since devices can be discarded or replaced [29].

To relate energy consumption to domestic activities, the characteristics of activities in private household should be clarified. Human activities are shaped by the habit of the human at home and can have the following features:

\paragraph{Arrangement} Activities in a home scenario can be concurrent, interrupted, interleaved or consecutive, also depending on the number of residents living in a home. 

\paragraph{Complexity} There are simple activities, e.g., doing laundry or more complex activities like Cooking or Working at home. Moreover, the appliance-activity mappings are non-exclusive due to the fact that sensor events can be mapped to one or more activities and vice versa [28].

\paragraph{Occurrence} Some activities are performed often and with an inflexible schedule like preparing dinner, others are rather rare and show a timely flexible character, e.g., listening to music. 

\paragraph{Diversity} The number and type of energy consuming appliances as well as the number and type of sensors are depending on the household since not every household has the same devices and sensors. Different appliances can also result in different possible activities. Furthermore due to differences in culture, region or character, there are different ways to perform the same activity which gives a particular contextual uncertainty [28]. 

\paragraph{Randomness} Despite the fact that humans have specific behavioral habits, activity patterns are not easy to extract since there exists a large randomness in activities of daily living due to the stochastic nature of human behavior [30].

Activity recognition models need to segment the stream of sensor events in a way to correctly map them with the actual activities that caused the specific sensor events. Due to this complex nature of activities, mapping sensor events to a label that indicates the corresponding activity is an ongoing challenge in the research of activity recognition [15]. 

Activity recognition contains methods that take various time-stamped sensorial data to extract patterns that describe different activities. Activity recognition is usually the assignment of an activity label to a sequence of sensor events that are generated from a smart home infrastructure [30]. Cameras and wearable sensors like gyroscopes and accelerometers are deployed in [31], [32] to get the necessary information. These sensors are intrusive and due to privacy issues people tend to not accept those devices in their private homes [33]. To understand body motions like sitting or walking also sensors in smart phones are used for activity recognition [34]. In a home scenario this approach is not very practical since residents do not carry their phone with them all the time which is a precondition to receive reliable data.  For activity recognition more non-intrusive, environmental sensors are often preferred such as motion, temperature, pressure or door sensors [35], [36].  For energy management purposes data from these sensors can be added to the energy consumption data on appliance level. 

Activity recognition methods are usually data-driven approaches. The data-driven methods take the information provided by sensors to build, infer and calibrate an activity recognition model [37]. They combine machine-learning techniques, data-mining methods and probabilistic models, such as Hidden Markov Models [38], [39], [40], Bayesian Networks [41] or Decision Trees [42], [43] to recognize activities by using inductive reasoning [44]. An activity recognition algorithm based on a decision tree model was presented by Lima et al [42]. The information from sensors and appliances were used to identify and categorize the activities. Six activity categories were determined each containing a set of activities performed in a certain time period. The sensors runtime can be compared with the average time of an activity stored in an activity time history table. The Euclidean distance measure is then used to determine the activity the recorded sensor runtime has the shortest distance to.

Most of the data-driven techniques are supervised needing labelled data for training. Marcello et al. [45] used a dataset consisting of various environmental sensor events labelled with the beginning and end time of a specific activity. 10 different activities were considered. Every sensor is treated as a feature. By implementing a Naïve Bayes Classifier every sensor is associated with a particular activity based on its distribution to be in a sensor sequence that is labelled with the name of that activity. The activity recognition was established by using a sensor-based observation window, sliding over and dividing the incoming stream of sensor events into segments. The classification is performed based on the probability of the sequence of sensor events to belong to a given activity. Applying a fixed sliding window requires a deep understanding of the given dataset since it can easily be too tight or too wide.

Alhamoud et al. [46] proposed a Random Forest as an activity recognition model using labelled data with the precondition that activities are only performed consecutively. The model received energy consumption data on appliance level, temperature, brightness and motion sensor readings as well as extracted temporal activity patterns as input. These patterns are the output of an a priori algorithm aiming at extracting temporal relationships between activities, e.g., A happens after B or before B. Due to the restrictions of only having consecutive activities, this approach is not very practical, since this scenario is far from being a real-world situation.

A three-stage framework to recognize the household’s activity and predict the next activity in a private household is presented by Du et al. [30]. Stage 1 consists of the definition of domestic activities, stage 2 of the activity recognition on the basis of weighted device usage data and stage 3 performs the activity prediction using an LSTM model.

Recognizing domestic activities by learning daily activities embeddings from sensor readings with a sequence-to-sequence model is introduced by Ghods et al [47]. The seq2seq model maps a sequence of sensor readings to itself to learn a set of features from raw sensor data. The learned embeddings were then utilized by a Random Forest classifier.
Acquiring a labelled and diverse training dataset is time consuming since it is usually manually done by the residents. Moreover the trained models are usually only specific to the trained subject in the same environment [48]. To bypass these limitations, there are also a few unsupervised approaches [49], [50] to solve the activity recognition problem but they usually lack in good performance results especially when it comes to detecting more complex activities.

Another approach to activity recognition is knowledge-driven. These techniques are based on contextual relationships that are described within an ontology and accordingly new instances are detected via deductive reasoning [37]. Ontologies have been generally used to describe a set of semantic concepts along with their relationship in a structural way. They tackle the challenge of knowledge-driven approaches by expressing knowledge in a structured, machine-understandable manner with the expressive power to support the deductive reasoning process [48] without requiring training. To recognize complex activities, data-driven techniques need sufficient training data which are usually hard and sometimes impossible to acquire [44]. On the other hand, ontologies suffer from interoperability and adaptation to different scenarios since they seem to be too generic and in static condition opposed to the dynamic nature of context data sources which are unknown in advance [37].

To overcome the limitations of both approaches, research started focusing on hybrid activity recognition by combining data-driven and knowledge-driven techniques to provide the necessities to handle uncertainty and also work in dynamic environments. Usually, ontological reasoning is used to detect complex activities since the recognition would require sufficient data to detect these activities via data-driven approaches. At the same time, it is more convenient to use data-driven reasoning to detect simpler activities than describing these activities in an ontology [44].

Stankovic et al. [28] presented a hybrid activity recognition approach by taking temporal energy consumption data on appliance level together with an activities ontology as input of their proposed activity inference algorithm [51]. Data for the ontology was acquired through semi-structured household interviews and video ethnography on technology ownership and usage. The ontology consists of a mapping of the relationship between appliances and activities distinguishing a direct or an indirect association of an appliance to an activity, e.g., the activity Working is directly associated to the computer and indirectly associated to the desk lamp. Reaching a practical implementation without having lots of sensors brings along the limitations through some uncertainty due to NILM’s possible misclassification if two appliances have a similar active power signature. Further limitations are introduced by confronting the stochastic nature of human behaviour to the static one of ontologies.

Ahmadi-Karvigh et al. [44] used data provided by power meters, motion and light sensors to detect actions defined as particular changes in the environment and, through a constructed ontology modelled by means of Description Logic (DL) language. Activities are recognized as combinations of different actions in an unsupervised approach. The user’s activity record was only collected for evaluation. However, there are limitations due to the required effort to initialize the system, e.g., for constructing the knowledgebase. 

Activity-aware systems for an efficient energy management are valuable since they can strongly adapt to the residents’ needs which is crucial to convince the users of using the system over a longer period. Thomas et al. [15] presented an activity-aware energy efficient automation system with the goal to turn off appliances that are not needed for the current activity and leave on devices which are required. More user adaptation is achieved by predicting the type and timing of the next activity within the next determined time window (10 minutes). The activity prediction problem in [15] is solved in the framework of imitation learning where the goal is to learn imitating an expert, that is demonstrating the desired behavior, in a way that generalizes to similar tasks and situations. 
 
Marcello et al. [14] pursued this approach by presenting a building energy management system for energy consumption reduction that is capable of controlling home appliances according to users preferences based on activity recognition and prediction using activity labeled sensorial data assigning an activity to one corresponding device. The activity prediction is performed for a specified time window (9 hours) starting from the assumption that activities are correlated with each other. An occurred activity brings along conditional probabilities for the possible following activity. During training of the activity prediction algorithm this is evaluated for every activity. 

\section{Methodology}

The suggested in this paper activity-based recommendation system is based on the multi-agent recommendation system of Riabchuk et al. [13]. In particular, our system aims at providing the residents with an activity schedule for demand response management in a comprehensible way being close to the daily private lifestyle with the focus of load shifting. Following this goal, a recommendation system can assume different shapes regarding the recommended time horizon, the precision of the recommended time windows, the number of activities and the included contextual information. Having the rational target to enable a practical implementation in private households, while ensuring minor user input, boundary conditions are determined for the shape of the proposed system.

The proposed recommendation system provides recommendation taking various aspects of user preferences into account. The system models user preferences depending on (1) the availability of the user, (2) energy costs and (3) CO2 emissions coming from the production of energy needed for domestic activities using electricity consuming devices. The user has also the possibility to state a ratio considering both, the costs of energy as well as the CO2 emissions. 
To factor in the user’s preferences, the system needs external day-ahead hourly electricity prices as well as hourly CO2 emissions predictions for the next day, historical energy consumption data on appliance level as well as an activity-device mapping in form of a vector representation. 
Based on this data the recommendation system provides activities’ beginning hours for the next 24 hours starting from the point in time the recommendations are provided by the system. The recommendations are made if the user’s availability is predicted by the system and the suggested activity schedule would reduce energy costs or CO2 emissions. All activities are included to provide the recommended activity schedule unless the user explicitly names activities to leave out of consideration. The technical requirements comprise smart plugs to get the energy consumption data on device level and an interface like a smartphone or a tablet for the user’s interaction with the system primarily to receive the recommendations.

\subsection{Framework}

The activity schedule suggested by the proposed recommendation system consists of the recommendation item, the shiftable activity, the recommended starting time and the predicted duration of the activity. The system provides a recommended activity schedule once a day at the same time for the next 24 hours starting from the specific point in time the recommendations are generated. The recommended activity schedule only considers activities the user specified during the implementation of the system. The list of activities can be updated while the recommendation system is already in use. The system produces recommendations per activity and specifies the activity starting times with hourly precision. The system provides as many starting times as the amount equal to the number of times the same activity is predicted to be conducted within the 24-hour recommendation horizon. The provided recommended activity schedule is displayed in the form presented in Table 1. 

\begin{figure}[h!]
	\centering
	\includegraphics[width=0.8\textwidth]{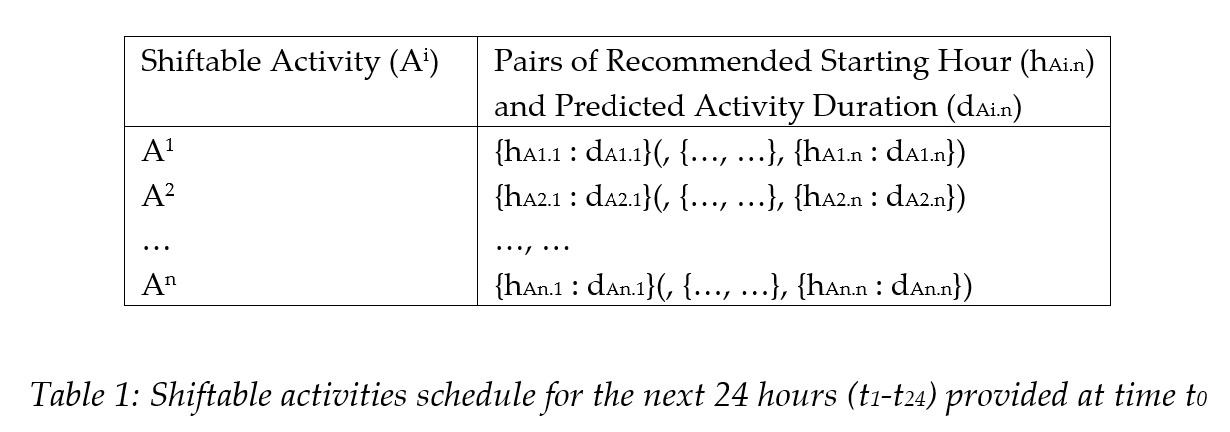}
	\label{}
\end{figure}

\subsection{Recommendation algorithm}

The proposed recommendation system extends the multi-agent architecture suggested by Riabchuk et al. [13]. The structure of our multi-agent recommendation system is shown in Figure 1. The Preprocessing Agent prepares the data for the Usage and the Availability Agent. The Usage Agent constitutes another preparation step for the Activity Agent whose output is combined with the outputs of the Availability, the Price as well as the CO2 Emissions Agent as the input for the Recommendation Agent that generates the recommendations described in the previous part. 

\begin{figure}[h!]
	\centering
	\includegraphics[width=0.8\textwidth]{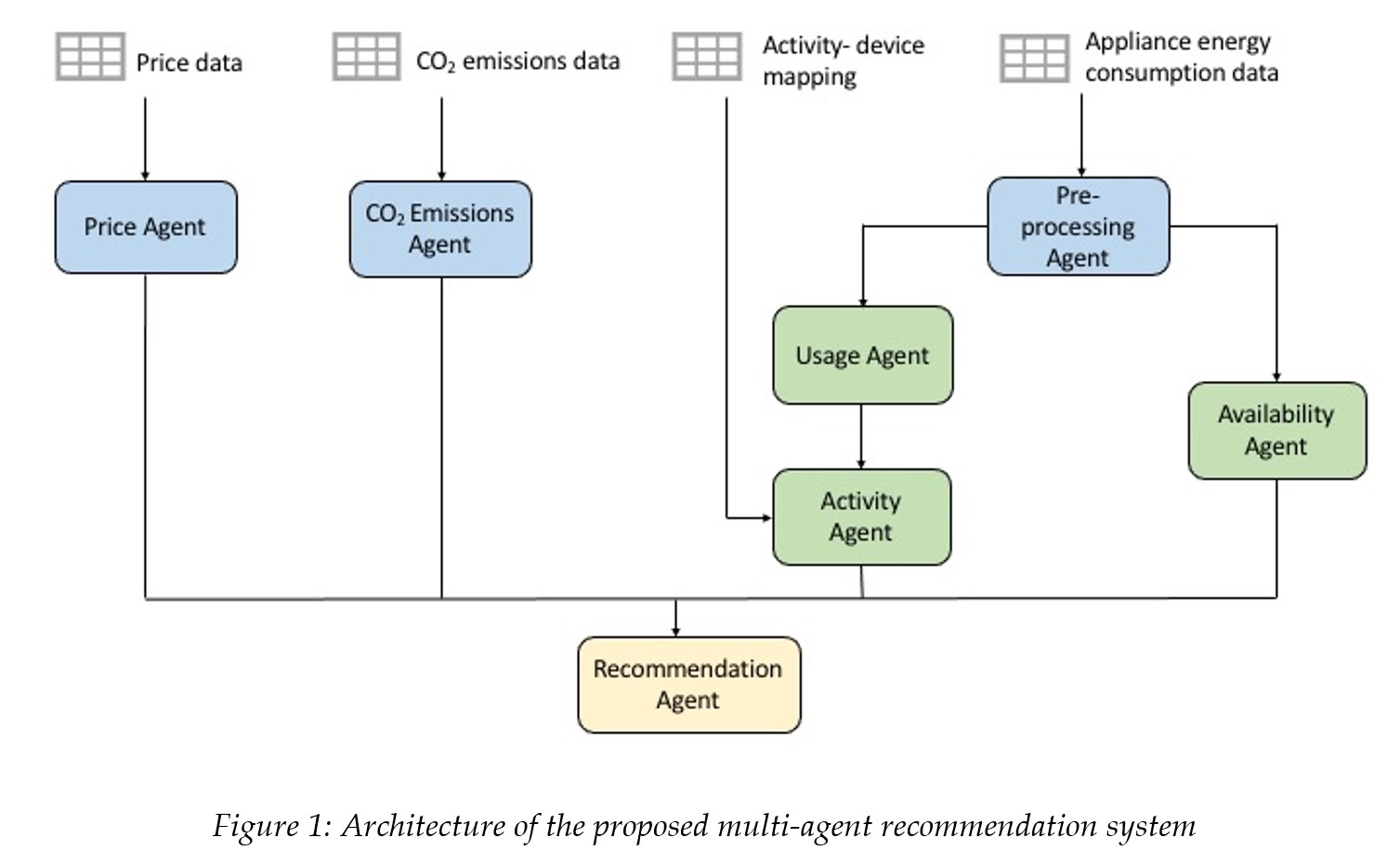}
	\label{}
\end{figure}

The \textbf{Price Agent} consists of a crawler module that is connected to a web page. The crawled web page contains the predicted trend of electricity prices on an hourly basis. The Price Agent selects these kinds of information for the next 24 hours and outputs a vector of electricity price signals for every hour of the 24-hour recommendation horizon $p_{24h} = (p_1, …, p_{24})_{24h}.$ The agent performs these steps every 24 hours on a specific point of time during the day.

The \textbf{CO2 Emissions Agent}, similar to the Price Agent, crawls hourly data indicating the predicted amount of CO2 emissions that will be produced during power production within the next 24 hours. It outputs a vector of CO2 emissions prediction for every hour of the 24-hour recommendation horizon $c_{24h} = (c_1, …, c_{24})_{24h}$. The agent performs these steps every 24 hours on a specific point of time during the day.

The \textbf{Preprocessing Agent} takes the energy consumption data on appliance level as input and performs several preprocessing steps on the data. The first step is to group the consumption data per appliance on an hourly level to align the data to the determined time window of the recommendations. Secondly, time features like the month, the day of week as well as the hour of the day are added to the dataset. 
Two time lag features are provided for the Availability Agent as one-hot encoded columns. Considering only appliances that indicate availability since a usage of these devices without the resident being at home is very unlikely, e.g., microwave, kettle, Hi-Fi or television, their hourly energy consumption is one-hot encoded and shifted by 1 hour and 1 week.
Equally, two time lag features are calculated for the Usage Agent specifying the usage of each appliance in a one-hot encoded representation shifted by 1 hour and 1 week. To provide the time lag features a threshold is used to indicate when an appliance is in use to prevent misinterpretations due to noise emerging from automatic actions. The Preprocessing Agent outputs a large dataset comprising the preprocessed hourly consumption data as well as features the following agents can choose from. 

The \textbf{Availability Agent} takes all time features as well as the time lag features referring to the availability as input with the goal to output probabilities indicating the user’s availability for every hour of the 24-hour recommendation horizon, in the form $(\pi_1^{AV},…,\pi_{24}^{AV})$. The using of time lag features transforms the time series forecasting problem into a supervised learning binary classification problem taking the original availability indicating one-hot encoded vector as the target variable. As a benchmark a logistic regression model is used trained on consumption data for one year to perform the prediction task. To test whether the prediction performance of the benchmark model can be improved, a random forest as well as a neural network are applied. A headstart of four weeks of data is used to train the models before they make their first prediction. The random forest and the neural network models are tuned every four months using the new energy consumption data cutting of the oldest two months of data to lead the focus of the model on the current energy consumption behaviour of the household.

The Usage Agent calculates usage probabilities for all devices that are being used for carrying out shiftable activities for every hour during the recommendation time, in the form of $(\pi_1^{DEV i},…,\pi_{24}^{DEV i})$. The time features as well as the time lag features referring to the usage of the individual appliances form the input of the agent. Due to the use of time features the problem is a binary classification problem with the original usage indicating one-hot encoded vector as target variable. A logistic regression as a benchmark, a random forest and a neural network were iteratively trained for every appliance using data of a time frame of one year. Similar to the implementation of the Availability Agent, the models are trained and tuned on data of four weeks before they make their first prediction and the random forest and the neural network models are tuned every four months on the new consumption data cutting of the oldest two months to let the models focus on the current behaviour of the household’s energy consumption. 

The \textbf{Activity Agent} provides activity probabilities for every shiftable activity during the 24-hour recommendation horizon, in the form of $(\pi_1^{ACT i},…,\pi_{24}^{ACT i})$. It applies a vector space model to calculate the similarity between the vectors constituting the hourly predicted appliance usage probability and the vector representation of each domestic activity of the private household against the backdrop of the energy consuming appliances used in those activities. 

The Activity Agent receives the output of the Usage Agent as well as a file containing the activity-device mappings in form of a vector representation as input. For each household an activity vector representation file has to be prepared that provides information on the relation of the household’s individual activities between the energy consuming appliances used to carry out the activities. 
The purpose of this vector representation is knowability, that is, to link measurable information on appliances to the set of activities characterising domestic life.

To compile the activity-device mapping in form of a vector representation, all devices that the recommendation system should take into consideration have to be mapped with activities that can be carried out with the considered devices. Ideally, the devices are mapped exclusively to one activity.  For instance, a washing machine definitely indicates the activity \textit{Laundering}, a dishwasher is an identifying device for the activity \textit{Cleaning}, a television indicates the activity \textit{Entertaining} and so on. The vector representing an activity consists of numbers of 0, no activity-device relation, or 1, identifying activity-device relation. Table 2 shows an example of an activity-device mapping in form of a vector representation for a fictional private household.

\begin{figure}[h!]
	\centering
	\includegraphics[width=0.8\textwidth]{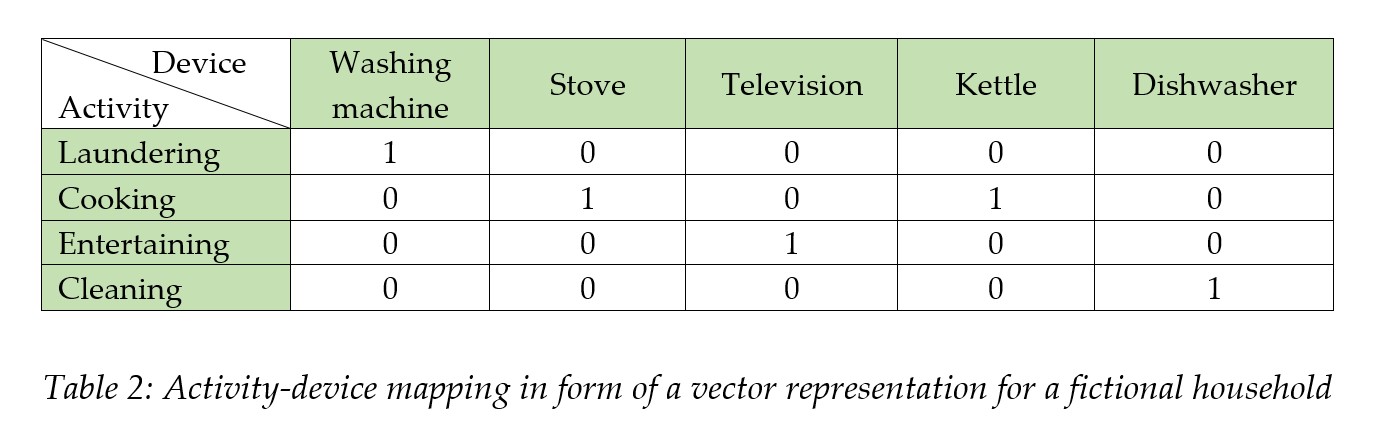}
	\label{}
\end{figure}

Five different appliances are considered and mapped to four different activities. Each appliance is mapped to one activity indicated by 1. As shown with the activity Cooking, an activity can have several devices that are mapped to it which is realistic. 
With this mapping the Activity Agent receives a vector representation for every activity $A^i$ in the form of $A^i = (R_{d_1}, R_{d_2}, …, R_{d_n})$ with $R_{d_n}$ specifying the relation $R$ in numerical form between device $d_n$ and $A^i$ as input for the vector space model.

Consequently, the vector space model has two input vectors each of the length equal to the number of devices $n_d$ considered by the recommendation system. Hence, the vector space has nd dimensions, one for each appliance ranging from 0, no device usage, to 1, definite device usage. The first vector in the vector space is the vector representing the device usage probabilities per hour (vector 1). The second vector is the activity-device mapping vector for an individual activity (vector 2).
To determine the similarity of these two vectors, the cosine similarity is used. This metric is particularly concerned with orientation and generally used to measure the distance when the magnitude of the vectors does not matter. The cosine similarity metric is used in positive space ranging from 0, no similarity, to 1, total similarity. The formula of the cosine similarity is as follows

\begin{figure}[h!]
	\centering
	\includegraphics[width=0.6\textwidth]{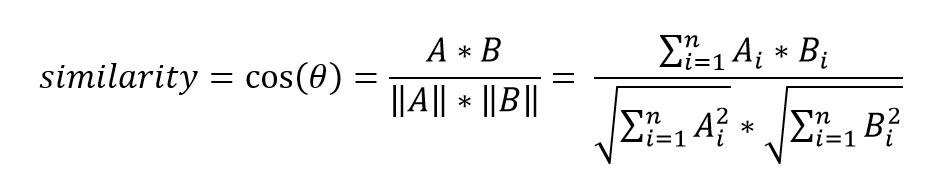}
	\label{}
\end{figure}

A and B are the two vectors whose similarity is calculated, n the number of vector components and i is the iterator. The Activity Agent takes vector 1 (=A) of a specific hour and compares it with vector 2 (=B) of each activity of the same specific hour. The result is one cosine similarity measurement for each activity of the private household per hour. The higher the cosine similarity measurement, the closer are the compared vectors due to a similar orientation specifying that the probable used devices indicated by vector 1 are similar to the devices that are used to carry out a certain activity indicated by vector 2. This procedure also enables the detection of concurrent activities. The cosine similarity of each activity vector is calculated and even though vector 1 states a high usage probability of identifying devices of different activities, both will be considered with a respective similarity measurement since the vector orientation will be closer to a vector indicating the usage of the identifying devices than to a vector not specifying such usage.

To provide the activity probabilities for every shiftable activity the computed cosine similarity calculations are transformed into probabilities by dividing each measurement by the sum of all measurements of that hour.

The \textbf{Recommendation Agent} receives the output of the Price Agent, the CO2 Emissions Agent, the Availability Agent and the Activity Agent to calculate an activity schedule for the next 24 hours. It also takes an availability threshold, specifying the minimum predicted probability for a user expected to be available, an activity threshold, specifying the minimum probability for an activity expected to be carried out, a ratio indicating to which extent the Recommendation Agent has to consider CO2 emissions over energy prices, a Boolean variable $aval\_off$ giving the system the information to consider availability hours for flexible activities or not and an average energy consumption per device.

At first the agent calculates the overall greenest and the overall cheapest hour of the day. It then computes the hours of availability depending on the availability threshold. The activity hours are determined depending on the activity threshold and stored in a dictionary with the keys being the predicted starting time and the value being the duration of the activity. 

Afterwards the Recommendation Agent iteratively looks at every possible activity to calculate the best starting hours as well as the emissions and price savings by multiplying the predicted duration of the activity with the sum of the average energy consumption of the devices mapped to the activity and being predicted to be used for the execution of the activity. To do so the activities are divided into groups referring to their flexibility to shift starting hours.

\textbf{Flexible activities} make reference to activities like Cleaning or Laundering that can easily be brought forward or postponed to later. Washing machines, tumble dryer or dishwasher usually can be programmed to an individual starting time, leaving the user’s availability unnecessary which can result in higher savings. Consequently, the user can specify leaving the predicted availability hours out of consideration regarding the flexible activities.

\textbf{Slightly flexible} activities have a certain potential to be brought forward or be postponed but not throughout the whole recommendation horizon. Entertaining activities like watching TV or listening to music are not tied to certain starting hours but periods during the day, e.g., entertaining activities are more often carried out in the evening hours. Therefore, the range of possible starting hours are set to 1 hour before to 4 hours after the predicted starting hour not aligning the range with the availability hours due to the fact that the prediction of the availability hours is only linked to the usage of energy consuming devices. However, a user can still be available without using an appliance. Since the predicted starting hours are availability hours, the range of possible starting hours are predicted availability hours or are close to a predicted availability hour, enabling the possibility that the user is available nonetheless. 

\textbf{Inflexible activities} refer to actions that are more or less bound to certain starting hours such as Cooking or Working. Cooking refer to basic needs, Working usually sticks to the mandatory working hours, hence the range of beginning hours is small. However, to also provide recommendations for inflexible activities the range of starting hours is set to 1 hour before and 2 hours after the predicted starting hour providing the user with possible saving potential and the opportunity to bring forward or postpone the energy consuming parts in these activities if possible. Equally to the management of slightly flexible activities, the range of possible starting hours is not aligned with the predicted availability hours for the same reasons. 

For each predicted activity starting hour the Recommendation Agent takes a time window with the length of the predicted activity duration and slides over the range of possible hours of beginnings adding the emissions and energy prices for the specific hours. According to the specified ratio the starting hour with the lowest emissions or the lowest costs is recommended by the system. In the end the Recommendation Agent provides information on the total emissions and cost savings achievable by implementing all provided recommendations. 

\subsection{Evaluation of the Recommendation System}

The motivation for the implementation of the proposed recommendation system for demand response focussing on load shifting emerges from the potential CO2 emissions and costs savings. The higher the possible savings generated by the recommendation system, the greater is the user’s willingness to change one’s own energy consumption behaviour and integrate the system in its daily living in the long run. Hence, in the following sections the performance of the proposed recommender system is evaluated following the performance evaluation framework suggested by Riabchuk et al. [13]. The total performance of the recommendation system depends on the performance of its single agents. Therefore, the performance of the individual agents is measured first. Afterwards, the challenge of the cold start problem is quantified by evaluating the agents’ performance over time. Last but not least, the potential CO2 emissions and cost savings of the overall recommendation system are analysed.

\subsubsection{Performance Evaluation of Individual Agents}

The agents that are considered in the performance evaluation of individual agents are the Availability Agent, the Usage Agent as well as the Activity Agent. Their performance over a one-year timeframe is evaluated. As described in section 3.2 the Availability Agent and the Usage Agent conduct a common binary classification task using supervised learning. Therefore, the Area Under the ROC-Curve is used (AUC score) to measure the performance of these agents. The AUC score is calculated by training models for every 24-hour recommendation window in the dataset. The first model is trained on the data of the first month of the dataset to predict the user availability or the device usage within the next 24-hour which represents the first recommendation window. This time window will then be added to the training set for the training of the second model which predictions refers to the second 24-hour recommendation window and so on. For every model an AUC score is calculated, which are combined by computing their mean to receive a single AUC score for the agents. 

The Activity Agent performs a less standard task using unsupervised learning. Research focusing on unsupervised machine learning to predict activities mostly measure their results by external evaluation or an activity record was collected for evaluation purposes [44]. Both could only be provided by the user. With the goal of minimal user input for a more convenient and practical implementation this is not an appropriate way. Therefore, a measure to evaluate this task is suggested without the need of ground truth data. 

The predictions of the Activity Agent are based on the device usage probabilities. If a device’s usage is pretty likely in a certain hour, the activity that has this device as its identifying device is also pretty likely to be carried out. Taking this principle every hour a set of devices $S_{dev}^i$ that are predicted with a probability higher than a certain usage threshold can be translated in a set of activities $S_{act}^i$ containing all activities, the devices in $S_{dev}^i$ are identifying devices for. As a result, $S_{act}^i$ can be used as a target variable to compare the predictions of the Activity Agent with. 

The output of the Activity Agent are activity probabilities for every possible activity per hour. An hourly set of predicted activities $S_{act\_pred}^i$ can be compiled by taking all activities higher than a certain activity threshold. To measure the performance of the Activity Agents both sets needs to be compared and checked for equality, meaning both containing the same activities. This procedure can be denoted in the following equations:

\begin{equation}
	ID_{act}= \{ dev \mid dev_{prob}=1 \}, 
\end{equation}

where $ID_{act}$ is the set of identifying devices for each possible activity of the private household. The values for $dev_{prob}$ are provided by the activity-device mapping vector where 1 represents an identifying relation between the activity and the device.

\begin{equation}
	S_{dev}^i= \{ dev \mid \pi^{dev} > use\_th \},
\end{equation}

where $S_{dev}^i$ describes the set of devices whose device usage probability of hour i is greater than the usage threshold $use_{th}$.

\begin{equation}
	S_{act}^i= \{ S_{act}  \mid dev \in S_{dev}^i  \wedge dev_{pred} \in ID_{act} \},	
\end{equation}

where $S_{act}^i$ is the set of activities that have an identifying activity-device relationship with the devices of $S_{dev}^i.$

\begin{equation}
	S_{act\_pred}^i = \{ act_{pred} \mid \pi^{act\_pred}  > act\_th \},
\end{equation}

where  $S_{act_pred}^i$ describes the set of activities that are predicted by the Activity Agent with a probability greater than the activity threshold $act_{th}$.

\begin{equation}
	EQUAL_{act} = \sum_{i=1}^n (S_{act}^i = S_{act\_pred}^i) / n,
\end{equation}

where $EQUAL_{act}$ is the ratio of the sum of equal activity sets per hour $i$ over $n$, with $n$ being the total number of hourly activity sets in the recommendation horizon, in this case 24. 
Since the Usage Agent performs a preprocessing step for the Activity Agent also their performance evaluation is obviously linked. Therefore, for each prediction from the different models trained to evaluate the Usage Agent, the performance of the Activity Agent is measured using the proposed $EQUAL_{(act)}$ measurement. To receive a single evaluation value for the Activity Agent, the different measurements are combined by calculating their mean. 

where $ EQUAL_{act}$ is the ratio of the sum of equal activity sets per hour i over n with n being the total number of hourly activity sets in the recommendation horizon; in this case 24. 
Since the Usage Agent performs a preprocessing step for the Activity Agent also their performance evaluation is obviously linked. Therefore, for each prediction from the different models trained to evaluate the Usage Agent, the performance of the Activity Agent is measured using the proposed $EQUAL_{act}$ measurement. To receive a single evaluation value for the Activity Agent, the different measurements are combined by calculating their mean. 

\subsubsection{Cold Start Problem Evaluation}

Every recommendation system is challenged by the so called cold start problem. Recommendation systems need data to be trained on to make appropriate recommendations. At the system’s start there is a lack in training data, hence, the recommendation system might not be able to provide proper recommendations [13].  The problem could be bypassed by requesting more user input in the beginning. However, this does not come along with the goal of minimal user input.

Therefore, the proposed recommendation system is evaluated in terms of its quality to overcome the cold start problem. As described in [13] the cold start problem is quantified by training the Availability Agent, the Usage Agent as well as the Activity Agent on varying length of the training dataset (one model for each training data length per agent) first. Their performance is then evaluated by how they improve with increasing datasets. To ensure comparability the performance of all models is evaluated using the same test data.

For the evaluation of the agents the scores introduced in section 3.3.1 are used to measure how fast the performance of the single agents converge to be higher than a specified sufficient performance score. To exactly measure how long the proposed recommendation system is challenged by the cold start problem the number of days is calculated each of the evaluated agents need to reach the specified tolerance interval. The cold start problem is tackled by the individual agents from the first day their evaluation score is higher than the specified score. For the recommendation system as a whole, the cold start problem is completed by the time the Availability Agent, the Usage Agent as well as the Activity Agent solved the cold start problem. 

\subsubsection{Performance Evaluation of the Recommendation System}

The overall goal of the proposed recommendation system is to enable CO2 emissions and cost savings for the user. Therefore, the performance of the recommendation system is evaluated by quantifying how much a user can save by employing it over a period of one year. To calculate the potential CO2 emissions and cost savings acceptable recommendations are generated by the proposed system and the CO2 emissions and energy costs for executing the activities are computed with and without following the recommendations. In this case, a recommendation is defined as acceptable if the availability of the user and the activity targets for the corresponding hour are given. 

The potential CO2 emissions savings are calculated by using the day-ahead CO2 emissions generated by the CO2 Emissions Agent, the possible energy cost savings are computed by using the day-ahead energy prices provided by the Price Agent, both based on the true energy consumption data of the appliances used for the predicted activities. 

The potential savings are dependent on the emissions ratio specified by the user. Hence, the possible savings are calculated assuming three emissions ratio scenarios:

\paragraph{1.	Emissions rate = 1:} The user’s priority is the saving of CO2 emissions. The proposed system generates recommendations focussing on reducing CO2 emissions only, which can result in higher energy costs compared to not following the recommendation.

\paragraph{2.	Emission rate = 0.5:} The user wants the recommendation system to consider the saving of CO2 emissions and energy costs equally.

\paragraph{3.	Emissions rate = 0.0:} The user’s priority is the saving of energy costs. The proposed system generates recommendations focussing on reducing energy costs only, which can result in higher CO2 emissions compared to not following the recommendation.

The potential savings are calculated for the complete 24-hour recommendation horizon, specifying the performance of the recommendation system by indicating the total savings as well as the relative savings at the household level.

\section{Empirical Results}

\subsection{Data}

For the proposed recommendation system, four different data inputs are required. To analyze and evaluate the proposed recommendation system the REFIT Electrical Load Measurements dataset [52] is used. The dataset contains cleaned electrical consumption data in Watts, timestamped and sampled at eight second intervals. The energy consumption data was collected from October 2013 to June 2015 in 20 households, for nine different devices each, in Loughborough, United Kingdom. To validate the findings, the evaluation steps described in section 3.3 are performed for household 1 to 5. 

To enable the provision of day-ahead carbon intensity forecasts occurring during energy production by the Carbon Intensity Agent, it accesses the database provided by the National Grid Electricity System Operator (ESO) [53]. In partnership with the Environmental Defense Fund Europe, the University of Oxford Department of Computer Science and the WWF, they developed the world’s first Carbon Intensity forecast. They released an API to provide an indicated trend of national and regional carbon intensity of the electricity system in Great Britain to enable consumers to optimize their energy consumption behavior to minimize CO2 emissions. 

The Price Agent provides the day-ahead energy prices through the access to the online database for industry day-ahead prices for the United Kingdom [54]. As stated by [13] these prices are not the actual consumption prices paid by the households, but they are a realistic proxy in the case of variable prices on the household level since higher industry prices are a result from higher market demand which also results in higher energy prices for households with variable tariffs. 
Both databases do not provide data for the timeframe the consumption data were collected. Therefore, the dates of energy consumption, carbon intensities and energy prices used for analysis match with their day and month, but not their year. Nevertheless, they indicate a realistic proxy for the evaluation of the recommendation system.
The Activity Agent requires the activity-device mapping in form of a vector representation, which has to be prepared for every household individually following the instructions described in section 3.2.

\subsection{Output of the System}

The result of the recommendation system is a recommendation for every shiftable activity described by the usage of the household’s devices. The system requires the following inputs: 
the day the recommendations are generated for; 
the availability threshold; 
the activity threshold; 
the Boolean value  $aval\_off$; 
the emissions ratio; 
average consumption values per device.
Only hours with a greater probability of user availability and activities with a higher probability to be performed on the day the recommendations are provided for are considered for recommendation. 

The output of the system starts with an overview of the greenest hour, meaning the hour with the fewest carbon intensity forecast of the day, and the cheapest hour of the day not considering predicted availability hours. The system then generates the best beginning hours combined with the predicted activities’ duration on the date of recommendation if both the availability flag and the activity flag are 0, meaning their predicted probability is higher than the respective threshold. The system concludes its output by providing the possible emissions and price savings for the recommended date in reference to the emissions ratio specified by the user that are achievable by implementing all provided recommendations. An exemplary output is shown in Table 3 with $aval\_off$ set to True.

\begin{figure}[h!]
	\centering
	\includegraphics[width=0.8\textwidth]{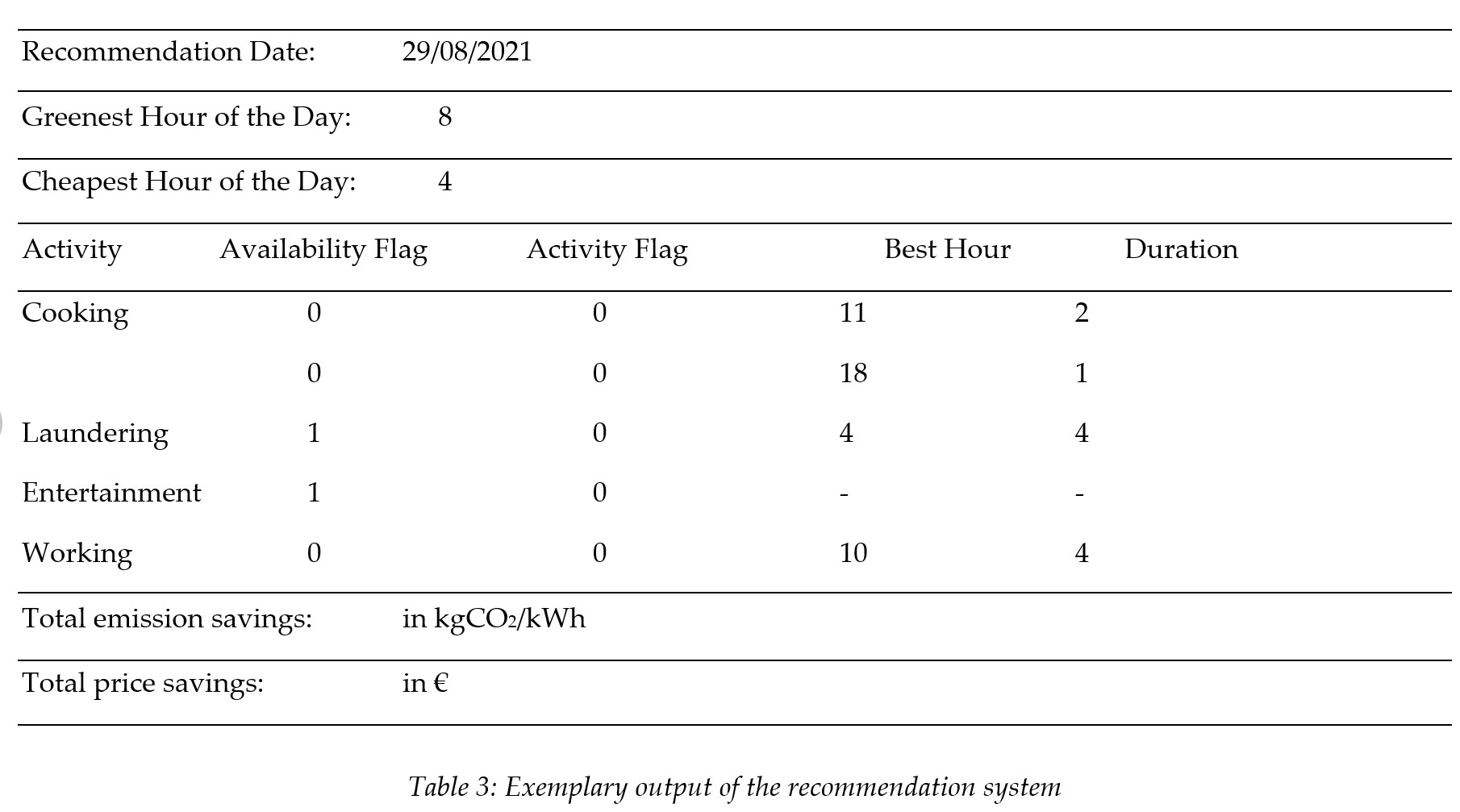}
	\label{}
\end{figure}

\subsection{Performance Evaluation}

\subsubsection{Performance of Individual Agents}

In order to evaluate the performance of the individual agents, the AUC score for the Availability Agent and the Usage Agent as well as the EQUAL score described in section 3.3.1 for the Activity Agent are calculated. 
To decide which model fits the best to the classification tasks of the Availability Agent and the Usage Agent, the consumption data of Household 5 with eight devices describing five activities was used to evaluate the three models. Table 4 shows the results of the performance evaluation. Table 9 in the Appendix A describes the type of devices for each household.

\begin{figure}[h!]
	\centering
	\includegraphics[width=0.8\textwidth]{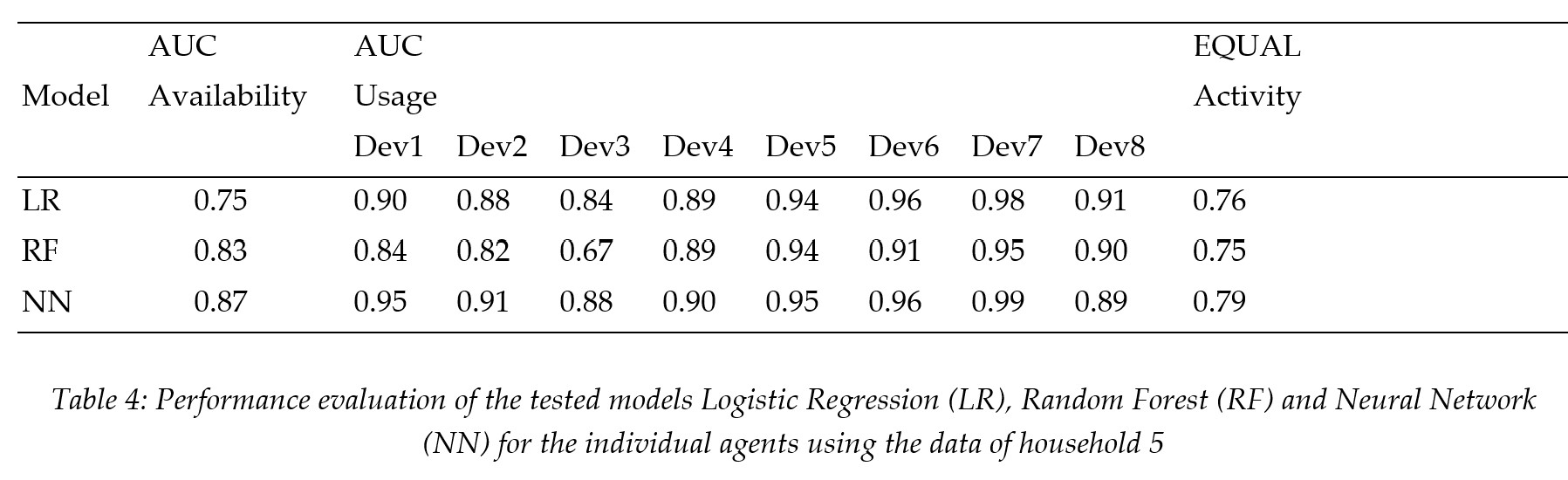}
	\label{}
\end{figure}


The neural network model generated the best AUC scores for both classification tasks. Since the output of the Usage Agent is the input of the Activity Agent, the performance of the Usage Agent influences the performance of the Activity Agent that also has the best results when the Usage Agent uses a neural network. Therefore, this model type is chosen for further performance evaluation. 
The performance results of the individual agents for the Households 1 to 5 with the Availability Agent and the Usage Agent using a neural network model are provided in Table 5.

\begin{figure}[h!]
	\centering
	\includegraphics[width=0.8\textwidth]{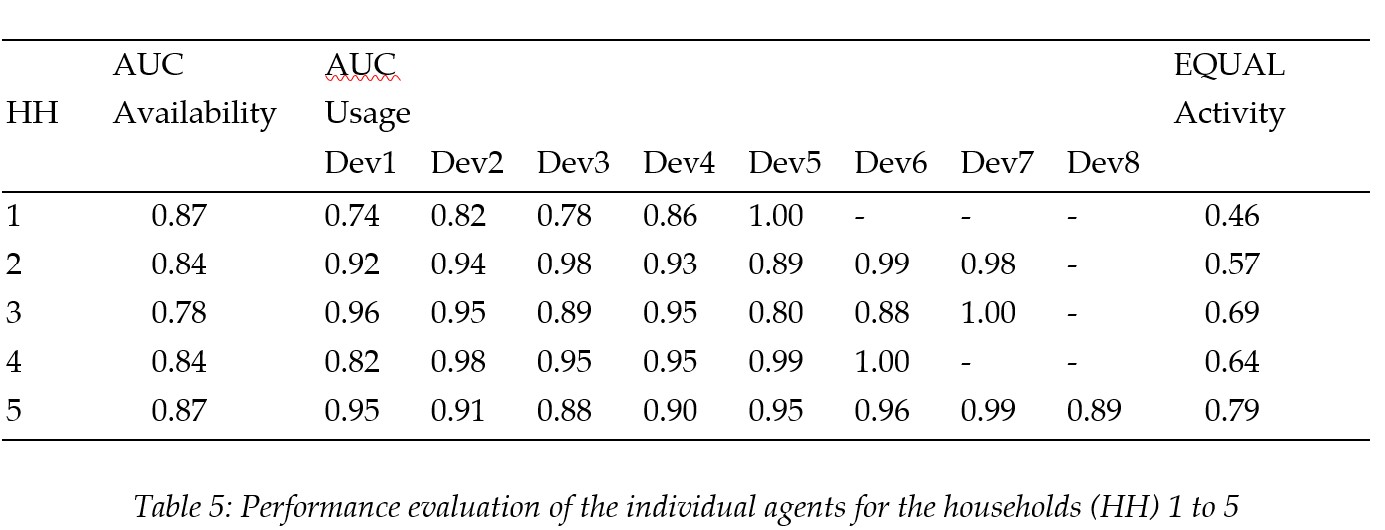}
	\label{}
\end{figure}


The performance quality of the Availability Agent and the Usage Agent could be validated for different consumption data of various households. There is only little variation of the Availability Agent’s and the Usage Agent’s performance being overall on a high level of AUC scores. There is a greater variance of the Availability Agent’s AUC scores for household 1 compared to household 3 as well as a difference of the performance of the Usage Agent for household 1 compared to the Usage Agent’s performance for household 2 to 5. A visual inspection of the energy consumption data of the households did not make a specific pattern evident. However, the prediction performance could possibly be influenced by a volatile behaviour of device usage during the day and over time. Moreover, the user’s availability also fluctuates across days and over time, especially in the case of weekend trips or longer vacations. 
The EQUAL score indicates how often the hourly set of activities given by the predicted devices through the activity-device mapping is equal to the hourly set of predicted activities higher than a certain probability. Therefore, the EQUAL score of household 5 shows that on average 19 hourly sets of 24 hourly sets were equal. 

The performance of the Activity Agent depends on the performance of the Usage Agent which explains the high score for household 3, 4 and 5 and the lower score for household 1. However, besides having good performance results of the Usage Agent for household 2, the performance of the Activity Agent does not absolutely reflect that. A reason for this might be the dependence of the EQUAL score calculation on thresholds. To compare the hourly activity sets, thresholds on the probabilities of device usage and the activities are necessary to filter activities out of the sets otherwise the sets would always be equal and the score would lose its expressiveness. An optimisation of the thresholds for the highest EQUAL score could be carried out which would be quite time consuming. Therefore, the EQUAL score forms an approximation for the quantification of an unsupervised learning problem. The EQUAL score of 0.57 indicates that on average for household 2 the Activity Agent could at least predict 11 hourly sets of 24 hourly sets correctly.

\subsubsection{Cold Start Problem}
In order to measure the impact of the cold start problem on the Availability Agent, the Usage Agent as well as the Activity Agent, the system’s performance over a year is analysed. As described in section 3.2, the neural network models of the Availability Agent and the Usage Agent are tuned and trained on 28 days to perform their first prediction on the 29th day. Therefore, the cold start problem is solved after 28 days at the earliest. During the evaluation process the models are then trained with different lengths of training data and tested on the same test data set. The models are tuned every 4 months with 6 months of consumption data, cutting of the oldest 2 months.

The cold start problem of the Availability Agent and the Usage Agent is solved if the performance of the models first hit an AUC of 0.79, the Activity Agent an EQUAL value of 0.65. A visualization of the models’ performance of the agents trained on energy consumption data of household 5 is shown in Figure 2. The models are tuned after around 90, 230 and 350 days, since the 28 days of tuning and training are not included in the graphs.

\begin{figure}[h!]
	\centering
	\includegraphics[width=0.8\textwidth]{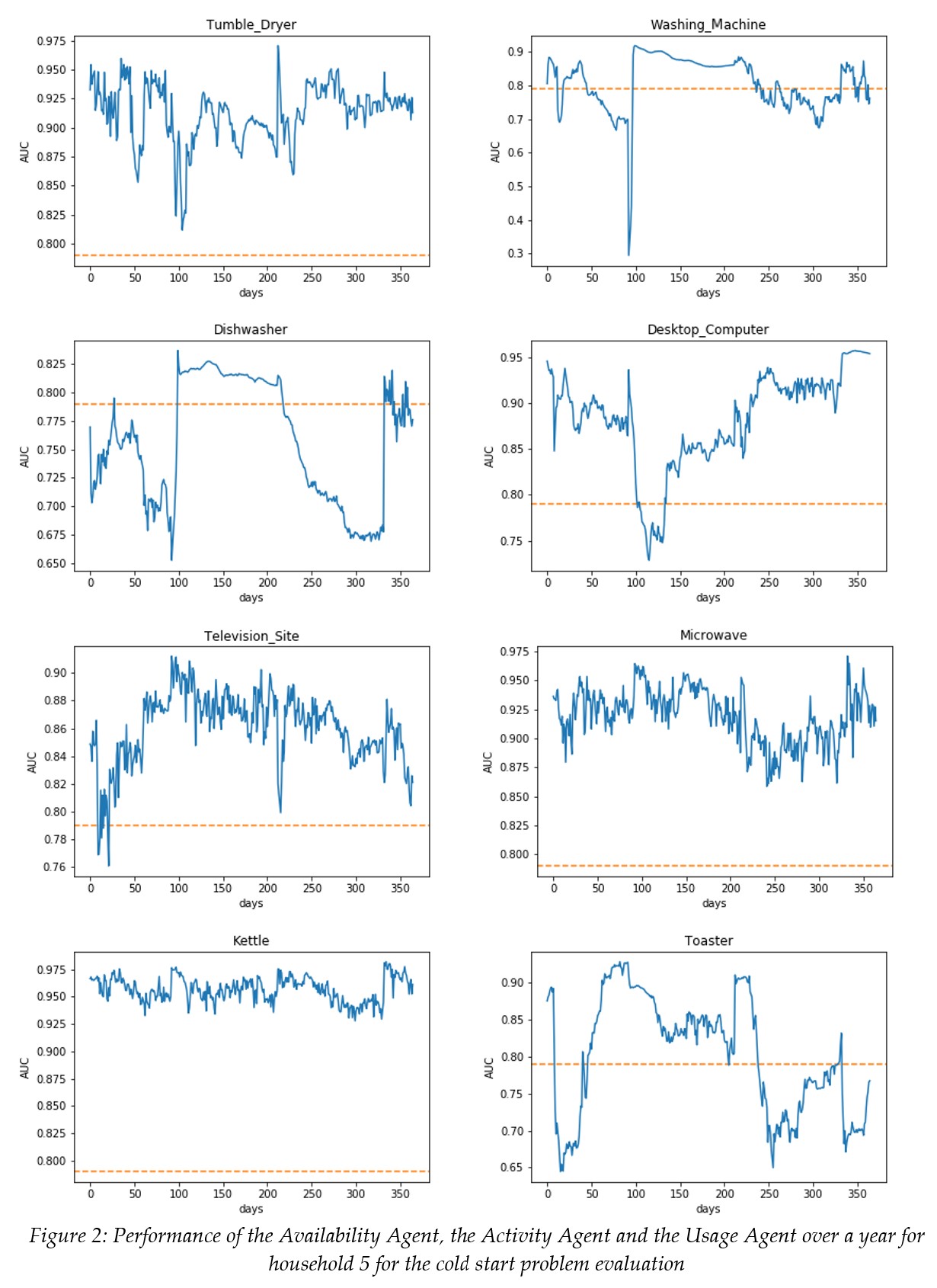}
	\label{}
\end{figure}


The Availability Agent reaches rapidly acceptable performance measures solving the cold start problem after about 2.5 months, which represents a quite constant period of time across different households (see Table 6). On the contrary the cold start scores of the Usage Agent vary across the different devices and households. While for some devices the cold start problem can instantly be solved after 28 days, the Usage Agent’s performance can overcome the cold start problem after 2 to 6 months or referring to Dev1 of household 1 cannot solve the cold start problem within a year of training. A reason for these measurements could be an infrequent usage of devices or a chaotic usage that is not following a pattern on a daily basis. Cutting off old usage data at every new tuning step is to make the model focus on the current device usage to learn new habits and unlearn old ones.  This way of model tuning can increase the models’ performance which, for instance, is shown in the graphs of the Washing Machine, the Dishwasher and the Desktop Computer after the tuning after 90 days. On the other hand, in some cases the model performances decreased after the tuning process which can be seen in the graphs of the Dishwasher and the Toaster after the tuning after 230 days. In this instance the model could not retain possible device usage changes. 

The cold start problem of the Activity Agent is solved after around 1 month which is also quite constant across the different households. The performance of the Activity Agent is linked to the performance of the Usage Agent. This explains the higher cold start score for the Activity Agent of household 1. 

\begin{figure}[h!]
	\centering
	\includegraphics[width=0.8\textwidth]{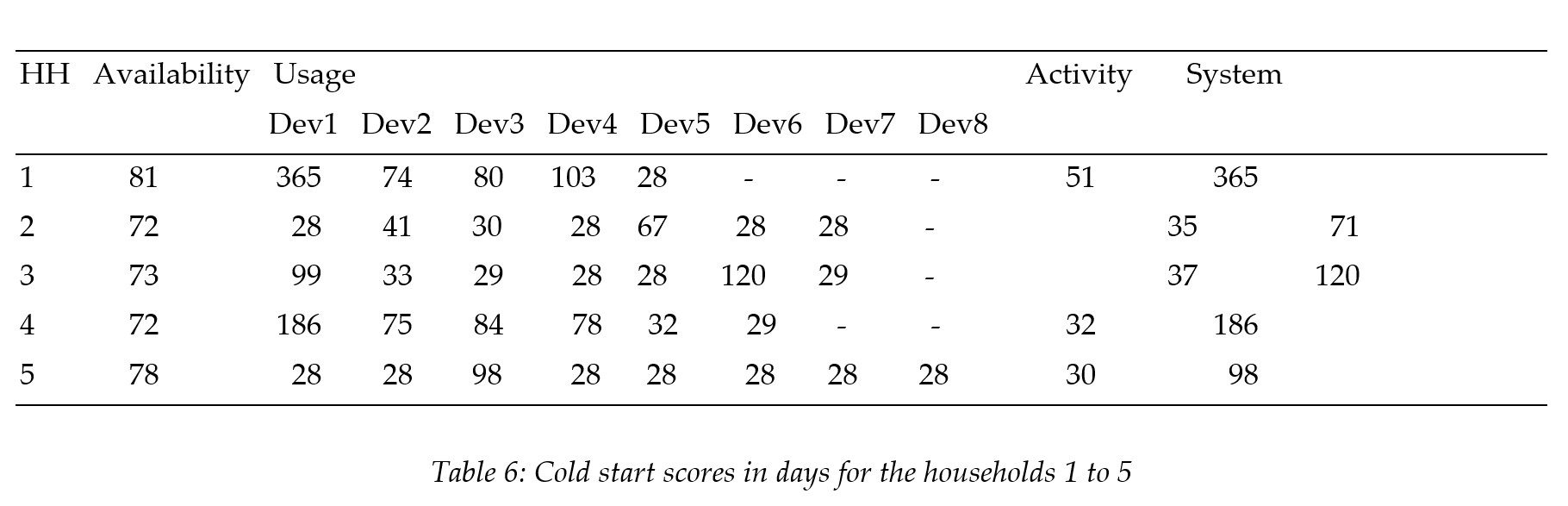}
	\label{}
\end{figure}

\subsubsection{Performance of the Recommendation System}

For the evaluation of the proposed recommendation system and its potential emissions and cost savings, the performance is measured using the following metrics at the household level: the number of recommendations per year, the number of recommendations per day, the total and relative emissions savings as well as the total and relative price savings. As described in section 4.1 the day-ahead emissions and price values used by the system to generate the recommendations and compute the total savings are not equal to the true savings a household would achieve applying the proposed system, but they form a useful proxy. However, for that reason the results are compared on the possible relative savings. 

For the generation of the recommendations, four hyperparameters need to be specified: $aval\_off$, emissions ratio, availability threshold and activity threshold. These are initiated by the system and can further be aligned by the user. For an optimal initialization, changes in the distribution of the recommended activity launching hours as well as in the performance measures due to changed hyperparameters values are analysed. Moreover, a grid search for the highest total savings on the household level is performed to confirm the best parameters found in the previous analysis. 

The first parameter $aval\_off$ specifies whether the system should consider the predicted hours of availability to generate recommendations for flexible activities. If this variable is set to true, the system can shift the launching hours of activities like Cleaning or Laundering to any hour of the recommended time horizon. It is reasonable that by setting this value to true more savings can be achieved since the range of hours to shift activities to is bigger and with this the possibility to change these activities to the greenest or the cheapest hour of the day even if this hour is not a predicted hour of availability. The performance analysis of the recommendation system shows that with the variable $aval\_off$ set to true relative savings increase by 50\%. Therefore, during the further presentation of the performance evaluation the $aval\_off$ variable is set to true, to show the full saving potential of the system.

The emissions ratio indicates to which extent the Recommendation Agent has to consider CO2 emissions over energy prices. An emissions ratio equal to 1.0 specifies a focus on emissions savings only, an emissions ratio set to 0.0 indicates a focus on price savings only, an emissions ratio equal to, for instance, 0.5 considers both, the amount of CO2 emissions and the energy price, equally. Figure 3 shows that the average day-ahead CO2 emissions per hour and the average day-ahead prices per hour are positive correlated. Therefore, a total focus on emissions savings can also lead to price savings and vice versa.

\begin{figure}[h!]
	\centering
	\includegraphics[width=0.8\textwidth]{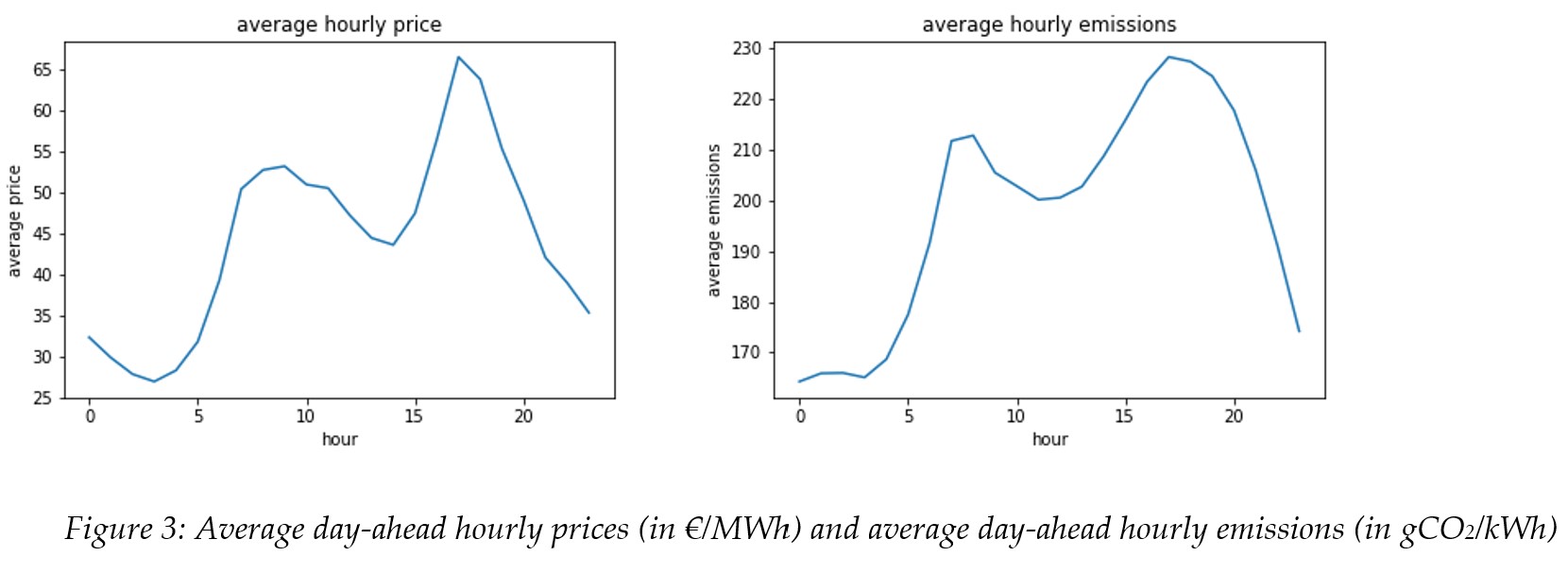}
	\label{}
\end{figure}

The availability threshold specifies the number of availability hours. Hence, the lower the threshold the higher the number of availability hours, the bigger the range of hours to shift activities to and generate savings. 
The activity threshold influences the number of predicted activities as well as the duration of these activities. More activities lead to greater possibilities of savings, longer activities on the other hand can also decrease possible savings.
To enable the highest possible savings the availability threshold is set to a constant value of 0.15 to analyse the influence of the activity threshold on the distribution of the activity starting hours and the performance measures over a year. Focussing on the savings of CO2 emissions (emissions ratio=1.0) the distribution of the recommendations over 24 hours shows the expected behaviour (see Figure 4). The starting hours are shifted to the hours with lower CO2 emissions, the various values of the activity threshold influence the number of recommendations and with this the total savings. The highest total savings are achieved with a low activity threshold. With increasing values for the activity threshold, the number of recommendations and with it the total savings are reduced. Moreover, the frequency of recommended starting hours around the noon hours decrease. Due to the fact that a higher activity threshold decreases not only the total number of activities but also the number of times a single activity is predicted per day explains this effect, mainly caused by flexible activities as they are first of all shifted to night or morning hours. 

Changing the focus of the recommendation to price savings (emissions ratio=0.0) with equal values for the remaining hyperparameters the distribution of the recommendation timings over 24 hours changes towards the hours with low energy costs as expected (see Figure 5). The distribution shows that activities are mainly shifted to low price night and morning hours even with a low activity threshold. The distribution remains as good as unchanged with an increase of the activity threshold.
Equally, a low activity threshold results in the highest price savings with the highest number of recommendations, that are both decreased with increasing values for the activity threshold. 

A further analysis of different values for the activity and the availability threshold shows that the highest possible savings are achieved with a low availability threshold combined with a low activity threshold. Higher values of the activity and the availability threshold lead to a reduction of the number of recommendations and with this a reduction of possible savings, but a similar distribution of recommendation timings (see Figure 8 and 9 in the Appendix B). 

\begin{figure}[h!]
	\centering
	\includegraphics[width=0.8\textwidth]{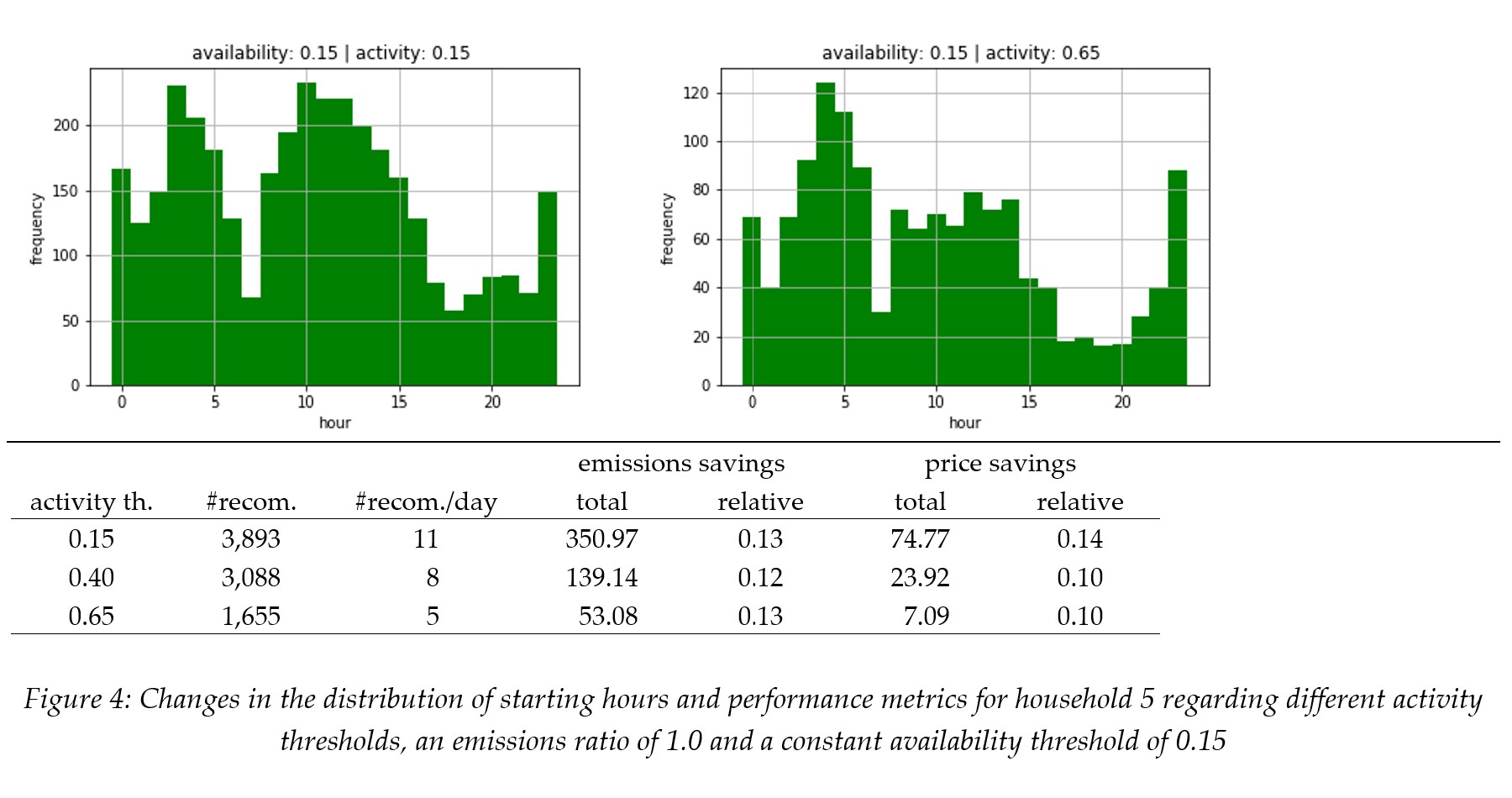}
	\label{}
\end{figure}

\begin{figure}[h!]
	\centering
	\includegraphics[width=0.8\textwidth]{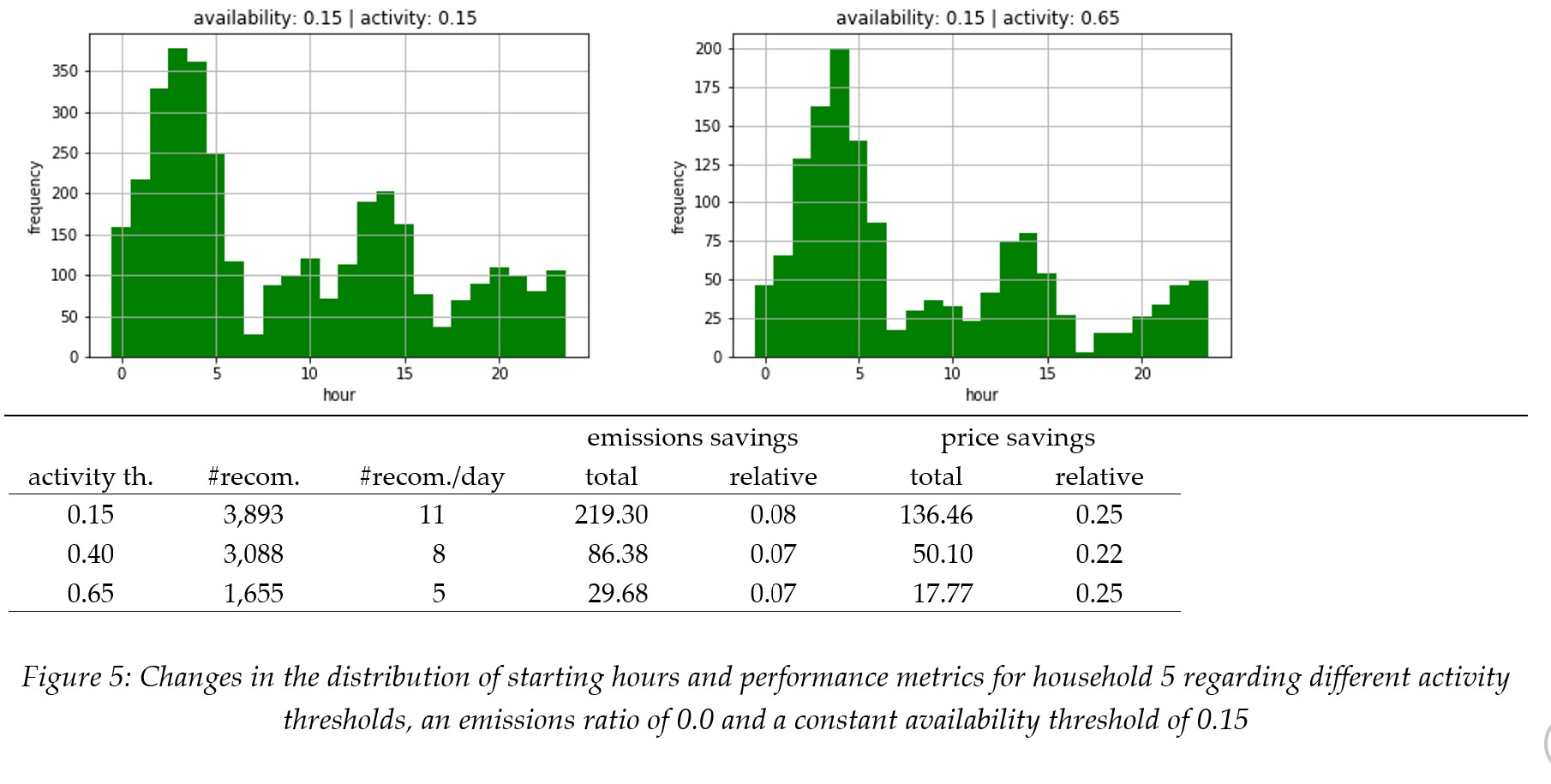}
	\label{}
\end{figure}

Setting the emissions ratio to focus on saving CO2 emissions comes with the price of lower or even negative price savings and vice versa. Setting the emissions ratio to 0.5 shows that the system considers both, emissions and price savings, while generating recommendations (see Table 7). The total emissions savings are lower than with an emissions ratio of 1.0, but higher than with an emissions ratio of 0.0, the total price savings are higher than with an emissions ratio of 1.0, but lower than with an emissions ratio of 0.0 which shows the expected behavior. With an emissions ratio of 0.5, sufficient emissions and price savings can be achieved.

\begin{figure}[h!]
	\centering
	\includegraphics[width=0.8\textwidth]{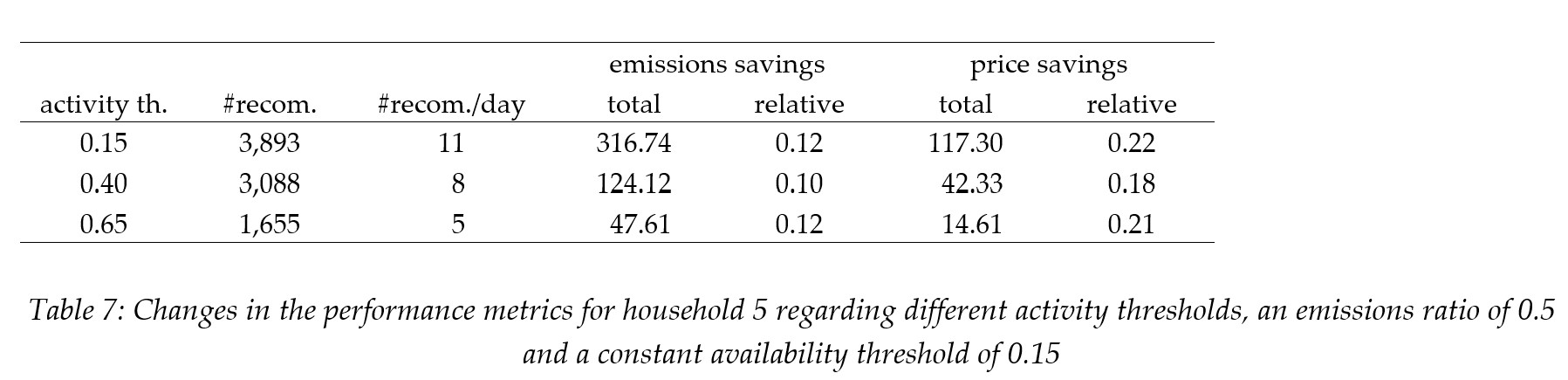}
	\label{}
\end{figure}

The distribution of the recommendation timings per activity show that as expected with an $aval\_off$ value set to True the highest saving potential comes with flexible activities. As Figure 6 shows, the starting hours for the activities Cleaning and Laundering are mainly concentrated in the low-emissions and low-price hours. Activities can be predicted more than one time per day, but each activity cannot be shifted to the same low-emissions or low-price hour. Hence, the first predicted flexible activity is shifted to the greenest or cheapest hours, the second one to the second greenest or cheapest hour and so on. This explains that, for instance, with a focus on price savings (emissions ratio = 0.0) some flexible activities are shifted to hours during the day which are not as low-price than hours during the night.

\begin{figure}[h!]
	\centering
	\includegraphics[width=0.8\textwidth]{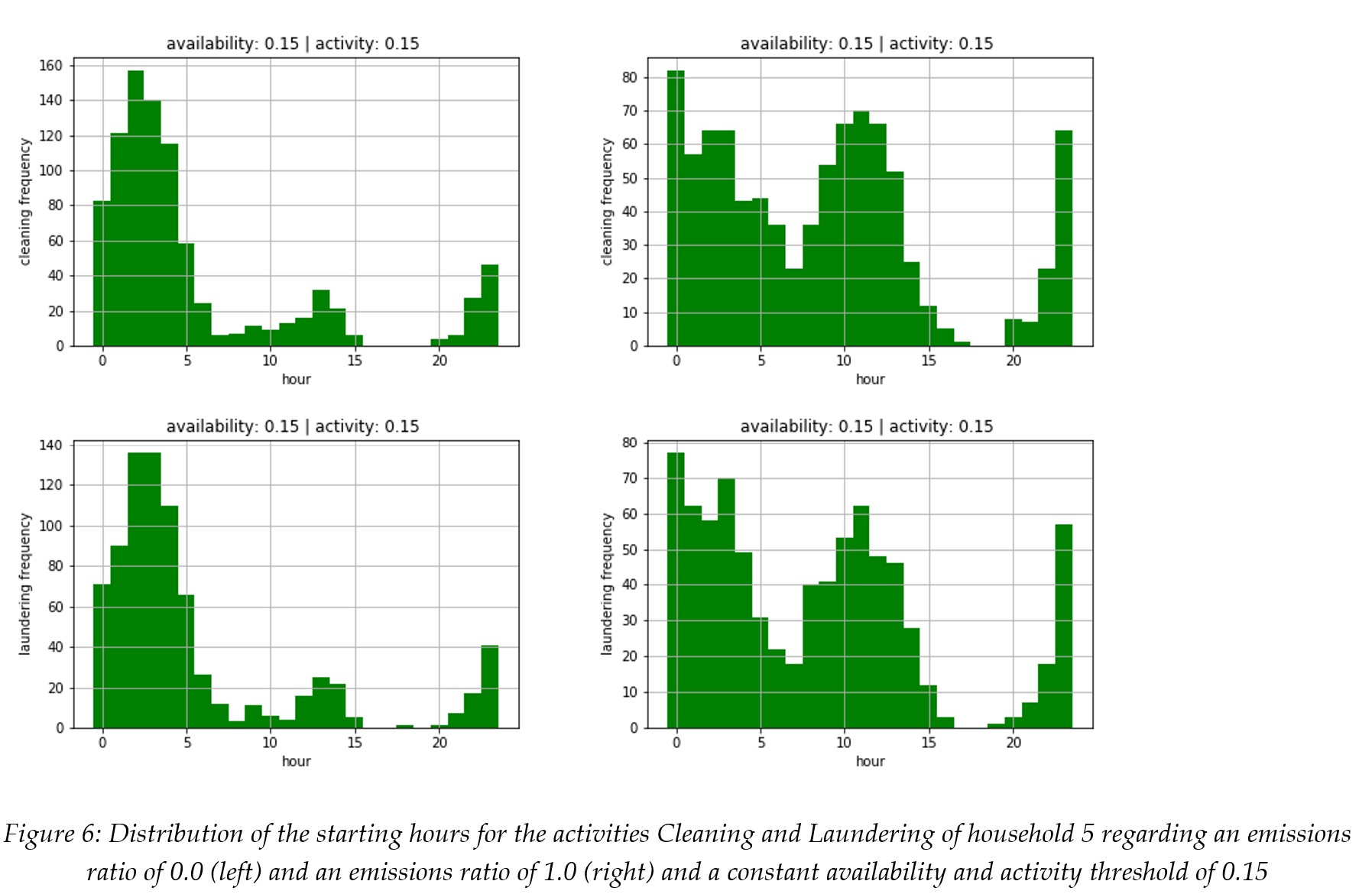}
	\label{}
\end{figure}

Slightly flexible or inflexible activities can be shifted only a limited number of hours which reduces the possibilities to save emissions or energy costs. The distribution of starting hours for such activities (see Figure 7) shows that the recommendation system provides recommendations to shift activities away from the peaks of hourly emissions and prices.

\begin{figure}[h!]
	\centering
	\includegraphics[width=0.8\textwidth]{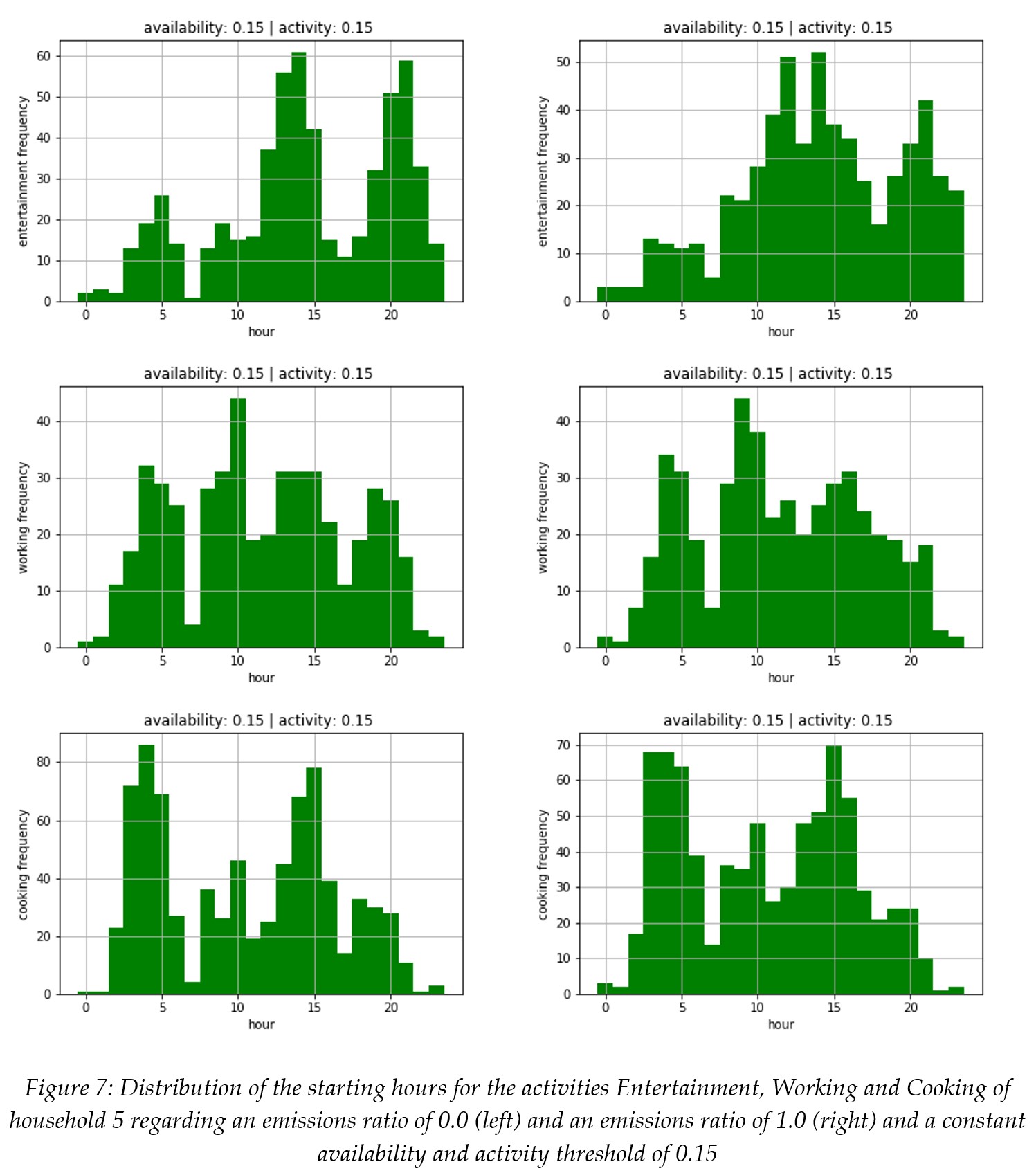}
	\label{}
\end{figure}

With regard to the performance measures across households (see Table 8) a grid search over the three hyperparameters $aval\_off$, availability threshold and activity threshold was performed on the household level, to (1) maximise the total emissions savings and (2) maximise the total energy price savings proofing the analysis that an $aval\_off$ value set to true and low availability and activity thresholds results in the highest total savings. Across the household an average of 12\% of emissions and an average of 7\% of energy cost savings can be achieved with a focus on emissions savings implementing all provided recommendations. Setting the focus on price savings, an average of 20\% of energy costs and an average of 6\% of CO2 emissions can be saved across the studied households implementing all recommendations.
The total emissions and cost savings vary across the different households since the amount of total savings is influenced by many factors like the amount of activities, the amount of devices used for the activities as well as the performance of the individual agents. The recommendation system for household 1, for instance, considers only 4 activities using 5 devices in total. Moreover, the performance of the individual agents for household 1 trails behind the performance of the agents for the other households. Another reason for the difference in total savings can also be the user behavior itself. If the user’s behavior carrying out the domestic activities, is already quite efficient without a recommendation system, the possible savings are obviously lower which could be an explanation for the smaller total savings of household 4.

\begin{figure}[h!]
	\centering
	\includegraphics[width=0.8\textwidth]{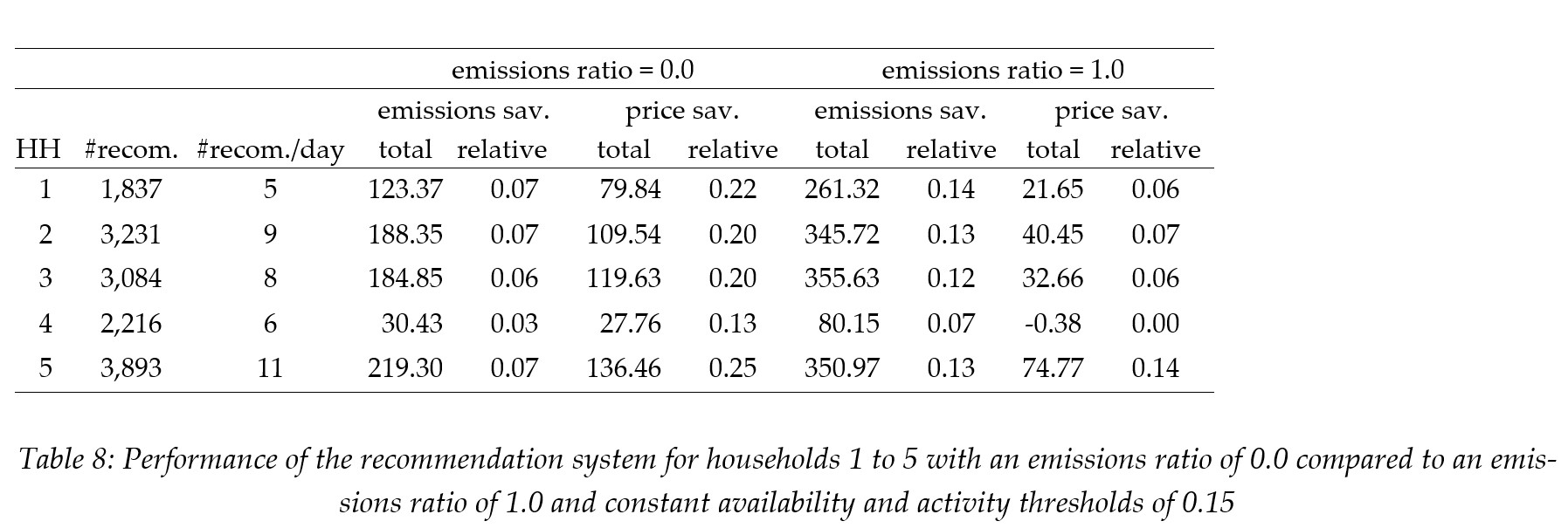}
	\label{}
\end{figure}

\section{Discussion}

The proposed activity-based approach to the recommendation system aims at a more efficient energy use regarding saving CO2 emissions emerging during energy production, reducing energy costs and preventing peak loads by providing recommendations to shift daily activities in private households. 

As the results show, deploying the proposed system across different households enables the possibility to save CO2 emissions and energy costs for the residents by shifting domestic activities. The amount of CO2 emissions and cost savings depend on the hyperparameters emissions ratio, $aval\_off$, availability threshold and activity threshold. The recommendation system is initialised based on the premise of achieving the highest possible savings, meaning with $aval\_off$ equal to true and the availability and activity thresholds set to a low value of 0.15. The value of the emissions ratio is initially set to 1.0 which enables high CO2 emissions, but also energy cost savings due to the positive correlation of the average hourly CO2 emissions and energy costs. This shows the potential of the system and the maximum possible emissions savings for the user which perhaps increases the motivation to change the energy consumption behaviour in the long term to exhaust all potential savings.

These hyperparameter initialisation results in the highest number of recommendations per day. The results indicate that with these settings an average of 8 daily recommendations are provided. In a household an average of 8 daily energy consuming activities is appropriate and recommendations to shift these number of activities should be realisable. In the end, the user is in control to implement all recommendations.

With every set of recommendations, the system also indicates which hour in the recommended time horizon is predicted to be the greenest and cheapest hour of the day independent of the predicted hours of availability. Providing these daily information rises awareness on the fact that the sources for the energy production as well as the energy price differ throughout the day and give some indication of the potential to control the ecological footprint and costs regarding the one’s own energy consumption. It is left to the user to use this information, for instance, to charge a laptop or a smartphone.


The recommendation system encourages changes in the energy consumption behaviour by providing recommendations to shift activities as a representative of the usage of various energy consuming devices. Producing recommendations using the same descriptive terms as the residents describe their life in their private households with, resonate more with the users, are more practical in use than following recommendations for the usage of single devices every day and benefit a long-term behavioural change towards an efficient energy consumption.

Applying the proposed system raises awareness that residents can have an impact on the CO2 emissions from the energy production and the energy costs with their daily energy consuming activities. In times of rising energy prices due to the energy transition, demand response initiated by end consumers brings back control over their own energy bill. The recommendation system aiming at saving CO2 emissions represents a valuable tool for the transformation to a renewable energy supply. Deployed by a great number of private households, the system can help decarbonise energy production to unburden the environment which is essential to mitigate climate change.

The energy consumption forecasts generated by the system can support balancing the power grid. Imbalances between the production and the consumption of electricity can result from forecast errors. If the energy production is too low, gas power plants are typically used as they are able to ramp up and down their electricity generation quickly, however, this type of energy production results in high CO2 emissions. Precise consumption forecasts facilitate to harmonize power production and consumption especially in the transition to a renewable energy supply. The recommendation system helps to flatten peak loads and to shift energy consumption to time of low load which assists grid stability. It supports the prevention of falling back to short-term energy production resulting in high emissions in times of high energy consumption and cutting of renewable energies like wind turbines in times of low energy consumption due to missing possibilities to store larger amounts of energy.

\subsection{Recommendations for the Practical Implementation}

The technical requirements to implement the proposed recommendation system are the availability of a smart meter as well as low-cost smart plugs to obtain the consumption data of every device used to carry out the shiftable activities. Jiménez-Bravo et al.[12] introduced a multi-agent recommendation system of a similar architecture using contextual data and being capable of obtaining consumption data using low-cost smart plugs. They tested the hardware as well as the data transmission by a specific protocol implemented through an application programming interface and confirmed the feasibility of the technical implementation of systems with similar architecture. 

Our system can be implemented within a smart home system or used via a smartphone application to receive the daily recommendations. A domestic device can easily be replaced as it has no influence on the activity itself, a new device requires a further smart plug and it has to either be added to an already existing activity, or a new activity has to be added to the activity-device mapping which can easily be adjusted. For every new activity the extent of flexibility (flexible, slightly flexible, inflexible) has to be determined. 

The recommendations are provided via an interface which offers the possibility to communicate with the system in order to change and adjust the initialised parameters. In case the user is not willing to accept a high number of recommendations per day, the user can increase the value of the activity threshold. This results in a reduction of the number of activities that can be shifted and with this, the number of recommendations per day with the drawback of decreased possible savings. The emissions ratio is initialised with the focus of the recommendation system to save CO2 emissions. This can also be adjusted by the user. If the reduction of energy costs is the prior motivation to deploy the system, the emissions ratio can be set to 0.0. If the user is indecisive, the emissions ratio can be set to 0.5. These parameters can be readjusted at any time in case the user’s perspective towards the number of daily recommendations or the motivation to deploy the system changes.

Applying the system to save energy costs requires reasonable day-ahead hourly price information and electricity providers to offer tariffs that pass the hourly price fluctuations to the customer. To use the system as a tool to save CO2 emissions, reliable hourly forecasts on the emissions that will emerge during the production of the energy that is provided by the user’s electricity provider have to be available. Moreover, to use the energy consumption forecast generated by the recommendation system to support the balance of the grid the household’s smart meter needs to be connected to a smart grid that enables the communication between the different players of the energy system for a secured energy supply on the basis of an efficient and reliable system operation [55].

\subsection{Contributions}

To provide utility-based recommendations for load shifting within demand response in private households, Jiménez-Bravo et al. [12] and Riabchuk et al. [13] introduce a recommendation system using a similar multi-agent architecture as well as contextual data and having the same hardware requirements. The proposed here recommendation system draws on the previous works by using simple hardware, the flexible architecture of a multi-agent system as well as a similar logic to generate recommendations by taking user preferences regarding device usage and user availability into account. However, the proposed system improves and enhances the previous findings in several aspects.

The recommendation systems introduced by Jiménez-Bravo et al.[12] as well as Riabchuk et al. [13] generate recommendations on when to use specific devices. We propose a system that produces recommendations to shift energy consumption based on daily activities. It “speaks” to the residents in the same descriptive terms as they would describe their domestic life. These familiar terms help to follow the recommendations more easily and with this, the user more likely sticks to the recommendations long enough to generate first savings which provides the necessary motivation to change the energy consumption behaviour in the long-term.

Thomas et al. [15] and Marcello et al. [14] present activity-aware systems, yet, requiring different sensorial data that is activity labelled. The proposed Activity Agent gets along with hourly energy consumption forecasts not needing any further sensorial data or activity labels. This reduces the implementation effort and costs and ensures minimal user input. Moreover, the introduced system extends the predicted time horizon compared to Thomas et al. [15] (10 minutes) and Marcello et at. [14] (9 hours) to a period of 24 hours. Marcello et al. [14] consider one appliance per activity whereas the proposed system also works with an amount of devices used for one activity which is more realistic.

Jiménez-Bravo et al. [12] generate recommendations for a device based on the historic usage frequency per weekday, Riabchuk et al. [13] provide recommendations based on the device usage probability forecasted for the recommended time horizon, both systems providing one recommendation per device per day maximum. Our proposed system also uses activity probabilities to provide recommendations, but on an hourly basis. Therefore, the recommendation system provides one recommendation for every time the probability for an activity to happen exceeds the set activity threshold in the recommended time horizon.  This reflects reality since domestic activities such as Cooking or Laundering can happen several times a day. Moreover, if the user is not capable or willing to follow one recommendation for a specific activity, the user still can follow the next recommendation(s) for that certain activity. Thereby, the user reduces possible savings but does not have to skip all possible savings regarding an activity.

Towards Jiménez-Bravo et al. [12], Riabchuk et al. [13] increase the precision of the usage costs calculation by determining devices’ average load profiles to estimate the costs. Our proposed algorithm outputs hourly probabilities of an activity to occur within the recommended time horizon. This enables to determine the activities’ duration by adding the numbers of consecutive hours that indicate a probability greater than the activity threshold. As a result, different duration per activity are used to calculate the emissions and energy cost savings which is closer to a realistic scenario. 

Riabchuk et al. [13] focus on shifting the usage of flexible devices such as washing machines or dishwasher, Jiménez-Bravo et al. [12] cover all shiftable devices but limiting the number of recommendations per device to not overload the user with too many recommendations. Recommending the shifting of activities for demand response reduces the number of daily recommendations on its own, since activities represent a specific number of devices that are used to carry out the activity. Instead of having several recommendations for single devices, they can be summed up by a recommendation for one activity the devices are used for.

The user availability is forecasted by predicting the device usage probability of those devices indicating an availability of the user. This is an important step to reduce the probability of generating useless recommendations to not deter the user from using the system. However, a user can still be at home without using any of the availability indicating devices which can result in a lower savings potential.  Our system faces this issue by providing the possibility to turn off the dependence on the predicted availability hours for recommendations for flexible activities by setting the parameter $aval\_off$ to true and also shifting the slightly flexible and inflexible activities predicted for hours of availability to hours that are not obliged to be a predicted hour of availability. 

The recommendation systems of Jiménez-Bravo et al. [12] and Riabchuk et al. [13] aim at saving energy costs with load shifting. The proposed system enhances the utility dimension by the possibility to set the focus on saving CO2 emissions or to allow the system to consider CO2 emissions and price savings with the shifting of domestic activities. Thereby, the system reaches a wider target group. In addition, the user can switch the focus of the system while using it. With this, the system offers a comfortable flexibility by the fact that the user does not have to stick to the one’s own decision the user made during the implementation of the system.

\subsection{Limitations}

A challenge in the development of recommendation system for demand response is the lack of useful datasets. The current system was originally intended to be created using German energy consumption data. The SMARTENERGY.KOM dataset collected by the Multimedia Communications Lab at the Technische Universität Darmstadt contains German energy consumption data of two smart home environments actually being labelled with nine domestic activities. The dataset was examined for a use in this work, but could not be utilised, due to the fact that the activities were performed consecutively only, which does not reflect a realistic daily life in private households especially in multi-person households.

Due to the lack of energy consumption data labelled with activities that fulfil realistic requirements the REFIT dataset was used in the end converting the activity prediction into an unsupervised problem. The evaluation of the activity prediction agent can therefore only be an approximation. The final evaluation resides with the user within a test of the system in a real environment.

The REFIT dataset contains energy consumption data of nine different devices per household with at least one being a non-shiftable one like a fridge or a freezer. Having only eight shiftable devices in a private household maximum, does not represent the reality. 
With the activities Cooking, Laundering, Cleaning, Working and Entertainment most of the devices’ usage can be described. However, the number of devices used to carry out an activity can differ. For instance, a predicted toaster usage leads to a prediction of the activity Cooking as well as the prediction of the parallel usage of the microwave, the kettle and the toaster. This misunderstanding can result in differences regarding the intended total savings if a different number of devices are used for an activity that the recommendation system calculated with. 

As already stated, taking user availability into account bears the risk of unused savings potential. Turning it off for flexible activities is up to the user and can be readjusted while the system is already in use. Shifting slightly flexible and inflexible activities to hours that are not obliged to be an hour of availability entails the risk to discourage users if the recommendations are not useful. The consideration of user availability faces the trade-off between not leaving savings potential and not deterring users from the system. However, due to the individuality of the users, setting a universal appropriate trade-off is exceptionally unlikely.

\subsection{Explainability}

The Availability Agent, the Usage Agent and the Activity Agent provide the basis for the Recommendation Agent to generate recommendations. Understanding how the individual agents compute their output is crucial for building trust in the findings of the agents and the overall working of the system. The Activity Agent uses a vector space model which is an algebraic model easily interpretable as the used formula is comprehensible. However, the output of the Activity Agent is dependent on the output of the Usage Agent, that is, also the trustworthiness of their results is linked. The Usage Agent as well as the Availability Agent use a neural network with one hidden layer. Although the applied neural network is not very deep, there is no insight on how the models approximate the function, that receives the input features to provide the output. Neural networks are so called black-box models lacking in transparency which can decrease credibility of the system applying the model.

As black-box machine learning models increasingly being applied to make important decisions, the demand for transparent deep learning models is increasing [56]. The explainable artificial intelligence is an emerging field in machine learning attempting to open up the black box to make the findings of non-linear programmed systems transparent. It offers a set of practical methods to explain the inner workings of deep learning models [57]. These methods include, for instance, the Layer-wise Relevance Propagation that operates by propagating the prediction backward in the neural network using a set of purposely designed local propagation rules to detect which input features the model uses to generate its prediction [58].  The Local Interpretable Model-agnostic Explanations technique is a further method to explain the classifiers for a specific single instance by manipulating the input data to create a series of artificial data containing only a part of the original attributes. Through the presence or absence of certain attributes their influence on the classification can be inferred [59]. Employing these methods to improve the explainability of the Availability Agent and the Usage Agent, increases traceability and trustworthiness in the overall recommendation system and the possibility to win customers being sceptical about the application of artificial intelligence.

\subsection{Future Work}

The proposed recommendation system can be additionally improved and enhanced in several directions.  The system strives for minimal user input. In the current version, the user can readjust the parameters of the system on its own to fit the one’s own needs. However, this could be automated by the system. For instance, the system could track which recommendations were accepted by comparing the predicted energy consumption considering all recommendations and the actual energy consumption. Moreover, the activity-device mapping has to be prepared manually before the start of the system covering all devices the system has to consider and assigning them to the activities the devices are used for. This process could potentially be automated by receiving the information through energy consumption data. The four weeks’ notice used to train the neural networks could be utilised to also retrieve the mapping. Hence, activity labels would be required which would have to be recorded by the user at the price of minimal user input.

The system provides recommendations for a 24-hour time horizon. Short-term deviations from the predicted activities the recommendations are based on cannot be considered. Providing also just-in-time recommendations could correct the previous ones. For instance, the system could detect a not predicted execution of the activity Entertainment and generate a just-in-time recommendation to shift the activity or provide information on how the execution of the activity reduces the predicted savings.

Due to the flexibility of the multi-agent architecture enhancements can easily be integrated into the framework of the recommendation system by adding or adapting agents without influencing the logic of the recommendations. Therewith, a weather agent could be added to provide weather forecasts that can influence the prediction of the user availability and activities especially on weekends. This agent could also be useful if the household possesses a photovoltaic system. With this, the system could also consider the availability of self-produced energy for the generation of the recommendations. Another agent could predict the usage of an electric car if available to store energy in times of low emissions, low costs or if the self-produced energy is not needed and then use the stored energy in times of high emissions, high energy costs or bad weather so that the solar system cannot produce energy.

\section{Conclusions}

This paper suggests an activity-based framework for the multi-agent recommendation system for a more efficient energy use in private households. Recommending activities for load shifting resonate more with the residents' domestic life, thus, fostering the acceptance and usage of the system over a longer period. The Activity Agent works with energy consumption data and does not require further sensorial data and activity labels. This reduces implementation costs, the need for extensive user input or  sharing of highly sensitive data. The system requires simple hardware and low computational resources. It assists households to save CO2 emissions and energy costs leaving the users in control. Furthermore, it supports grid operators by flattening peak-loads and stabilizing the grid.

\section*{References}

1.	NASA. Responding to Climate Change 2020.  20.10.2021. Available from: \url{https://climate.nasa.gov/solutions/adaptation-mitigation/}.

2.	European Commission. 2050 long-term strategy. 2020.  20.10.2021. Available from: \url{https://ec.europa.eu/clima/policies/strategies/2050_en.}

3.	Institute for Advanced Sustainability Studies. Long-term climate goals - Decarbonisation, carbon neutrality, and climate neutrality. 2015  20.10.2021]; Available from: \url{https://www.iass-potsdam.de/sites/default/files/files/policy_brief_decarbonisation.pdf.}

4.	myclimate. What are "negative emissions"? 2019  20.10.2021. \url{Available from: https://www.myclimate.org/information/faq/faq-detail/what-are-negative-emissions/.}

5.	European Commission. Clean Energy. 2020  20.10.2021]; Available from: \url{https://ec.europa.eu/international-partnerships/sdg/clean-energy_en.}

6.	Allouhi, A., et al., Energy consumption and efficiency in buildings: current status and future trends. Journal of Cleaner production, 2015. 109: p. 118-130.

7.	Guerra, O.J., Beyond short-duration energy storage. Nature Energy, 2021. 6(5): p. 460-461.

8.	Law, Y.W., et al. Demand response architectures and load management algorithms for energy-efficient power grids: a survey. in 2012 Seventh International Conference on Knowledge, Information and Creativity Support Systems. 2012. IEEE.

9.	Marinakis, V. and H. Doukas, An advanced IoT-based system for intelligent energy management in buildings. Sensors, 2018. 18(2): p. 610.

10.	Khalid, R., et al., Fuzzy energy management controller and scheduler for smart homes. Sustainable Computing: Informatics and Systems, 2019. 21: p. 103-118.

11.	Fioretto, F., W. Yeoh, and E. Pontelli. A multiagent system approach to scheduling devices in smart homes. in Workshops at the Thirty-First AAAI Conference on Artificial Intelligence. 2017.

12.	Jiménez-Bravo, D.M., et al., Multi-agent recommendation system for electrical energy optimization and cost saving in smart homes. Energies, 2019. 12(7): p. 1317.

13.	Riabchuk, V., Hagel, L., Germaine, F., \& Zharova, A. (2022). Utility-Based Context-Aware Multi-Agent Recommendation System for Energy Efficiency in Residential Buildings. arXiv preprint. arXiv:2205.02704.

14.	Marcello, F. and V. Pilloni, Smart Building Energy and Comfort Management Based on Sensor Activity Recognition and Prediction. Sensors, 2020. 1: p. s2.

15.	Thomas, B.L. and D.J. Cook, Activity-aware energy-efficient automation of smart buildings. Energies, 2016. 9(8): p. 624.

16.	Jordehi, A.R., Optimisation of demand response in electric power systems, a review. Renewable and sustainable energy reviews, 2019. 103: p. 308-319.

17.	U.S. Departement of Energy. Demand Response - Policy. 2021  07.09.2021]; Available from: \url{https://www.energy.gov/oe/services/electricity-policy-coordination-and-implementation/state-and-regional-policy-assistanc-4.}

18.	Singh, A., S. Doolla, and R. Banerjee, Demand side management. Encyclopedia of sustainable technologies, 2017: p. 487-496.

19.	verbraucherzentrale. Smart Meter: Die neuen Stromzähler kommen. 2021  09.09.2021]; Available from: \url{https://www.verbraucherzentrale.de/wissen/energie/preise-tarife-anbieterwechsel/smart-meter-die-neuen-stromzaehler-kommen-13275.}

20.	Logenthiran, T., D. Srinivasan, and T.Z. Shun, Demand side management in smart grid using heuristic optimization. IEEE transactions on smart grid, 2012. 3(3): p. 1244-1252.

21.	Curtis, J., et al., Why do preferences for electricity services differ? Domestic appliance curtailment contracts in Ireland. Energy Research \& Social Science, 2020. 69.

22.	Leitao, J., et al., A survey on home energy management. IEEE Access, 2020. 8: p. 5699-5722.

23.	Moreira, G. Recommender Systems in Python 101. 2020  12.09.2021. Available from: \url{https://www.kaggle.com/gspmoreira/recommender-systems-in-python-101.}

24.	Shaikh, P.H., et al., A review on optimized control systems for building energy and comfort management of smart sustainable buildings. Renewable and Sustainable Energy Reviews, 2014. 34: p. 409-429.

25.	Li, S., et al., A real-time electricity scheduling for residential home energy management. IEEE Internet of Things Journal, 2018. 6(2): p. 2602-2611.

26.	Gelvez García, N.Y., A.D. Ballén Duarte, and H.E. Espitia Cuchango, Multi-Agent System Used for Recommendation of Historical and Cultural Memories. Tecciencia, 2019. 14(26): p. 43-52.

27.	Katzeff, C. and J. Wangel, Social practices, households, and design in the smart grid, in ICT innovations for sustainability. 2015, Springer. p. 351-365.

28.	Stankovic, L., et al., Measuring the energy intensity of domestic activities from smart meter data. Applied Energy, 2016. 183: p. 1565-1580.

29.	Schwartz, T., et al., What people do with consumption feedback: a long-term living lab study of a home energy management system. Interacting with Computers, 2015. 27(6): p. 551-576.

30.	Du, Y., Y. Lim, and Y. Tan, A novel human activity recognition and prediction in smart home based on interaction. Sensors, 2019. 19(20): p. 4474.

31.	Kumrai, T., et al. Human Activity Recognition with Deep Reinforcement Learning using the Camera of a Mobile Robot. in 2020 IEEE International Conference on Pervasive Computing and Communications (PerCom). 2020. IEEE.

32.	Dua, N., S.N. Singh, and V.B. Semwal, Multi-input CNN-GRU based human activity recognition using wearable sensors. Computing, 2021: p. 1-18.

33.	Nitti, M., et al., When social networks meet D2D communications: A survey. Sensors, 2019. 19(2): p. 396.

34.	Ahmed, N., J.I. Rafiq, and M.R. Islam, Enhanced human activity recognition based on smartphone sensor data using hybrid feature selection model. Sensors, 2020. 20(1): p. 317.

35.	Liu, Y., et al., From action to activity: sensor-based activity recognition. Neurocomputing, 2016. 181: p. 108-115.

36.	Golestani, N. and M. Moghaddam, Human activity recognition using magnetic induction-based motion signals and deep recurrent neural networks. Nature communications, 2020. 11(1): p. 1-11.

37.	Rodríguez, N.D., et al., A survey on ontologies for human behavior recognition. ACM Computing Surveys (CSUR), 2014. 46(4): p. 1-33.

38.	Viard, K., et al. An event-based approach for discovering activities of daily living by hidden Markov models. in 2016 15th International Conference on Ubiquitous Computing and Communications and 2016 International Symposium on Cyberspace and Security (IUCC-CSS). 2016. IEEE.

39.	Asghari, P., E. Soleimani, and E. Nazerfard, Online human activity recognition employing hierarchical hidden Markov models. Journal of Ambient Intelligence and Humanized Computing, 2020. 11(3): p. 1141-1152.

40.	Müller, M., F. Biedenbach, and J. Reinhard, Development of an integrated simulation model for load and mobility profiles of private households. Energies, 2020. 13(15): p. 3843.

41.	Liu, L., et al., Towards complex activity recognition using a Bayesian network-based probabilistic generative framework. Pattern Recognition, 2017. 68: p. 295-309.

42.	Lima, W.S., et al. User activity recognition for energy saving in smart home environment. in 2015 IEEE Symposium on Computers and Communication (ISCC). 2015. IEEE.

43.	De Leonardis, G., et al. Human Activity Recognition by Wearable Sensors: Comparison of different classifiers for real-time applications. in 2018 IEEE International Symposium on Medical Measurements and Applications (MeMeA). 2018. IEEE.

44.	Ahmadi-Karvigh, S., et al., Real-time activity recognition for energy efficiency in buildings. Applied energy, 2018. 211: p. 146-160.

45.	Marcello, F., V. Pilloni, and D. Giusto, Sensor-based early activity recognition inside buildings to support energy and comfort management systems. Energies, 2019. 12(13): p. 2631.

46.	Alhamoud, A., et al. Extracting human behavior patterns from appliance-level power consumption data. in European Conference on Wireless Sensor Networks. 2015. Springer.

47.	Ghods, A. and D.J. Cook, Activity2vec: Learning adl embeddings from sensor data with a sequence-to-sequence model. arXiv preprint arXiv:1907.05597, 2019.

48.	Sukor, A.S.A., et al., A hybrid approach of knowledge-driven and data-driven reasoning for activity recognition in smart homes. Journal of Intelligent \& Fuzzy Systems, 2019. 36(5): p. 4177-4188.

49.	Li, F. and S. Dustdar, Incorporating Unsupervised Learning in Activity Recognition. Activity Context Representation, 2011. 11: p. 04.

50.	Trabelsi, D., et al., An unsupervised approach for automatic activity recognition based on hidden Markov model regression. IEEE Transactions on automation science and engineering, 2013. 10(3): p. 829-835.

51.	Liao, J., L. Stankovic, and V. Stankovic. Detecting household activity patterns from smart meter data. in 2014 International Conference on Intelligent Environments. 2014. IEEE.

52.	Murray, D., L. Stankovic, and V. Stankovic, An electrical load measurements dataset of United Kingdom households from a two-year longitudinal study. Scientific data, 2017. 4(1): p. 1-12.

53.	National Grid Great Britain. Carbon Intensity API [online database]. Available from: \url{https://www.carbonintensity.org.uk/.}

54.	entsoe Transparency Platform. Day-ahead prices [online database]. Available from: \url{https://transparency.entsoe.eu/transmission-domain/r2/dayAheadPrices/show.}

55.	Bundesministerium für Wirtschaft und Energie. Intelligente Netze. 2021  16.10.2021]; Available from: \url{https://www.bmwi.de/Redaktion/DE/Artikel/Energie/intelligente-netze.html.}

56.	Arrieta, A.B., et al., Explainable Artificial Intelligence (XAI): Concepts, taxonomies, opportunities and challenges toward responsible AI. Information Fusion, 2020. 58: p. 82-115.

57.	Das, A. and P. Rad, Opportunities and challenges in explainable artificial intelligence (xai): A survey. arXiv preprint arXiv:2006.11371, 2020.

58.	Montavon, G., et al., Layer-wise relevance propagation: an overview. Explainable AI: interpreting, explaining and visualizing deep learning, 2019: p. 193-209.

59.	Dieber, J. and S. Kirrane, Why model why? Assessing the strengths and limitations of LIME. arXiv preprint arXiv:2012.00093, 2020.

\newpage

\section*{Appendix}

\begin{figure}[h!]
	\centering
	\includegraphics[width=0.8\textwidth]{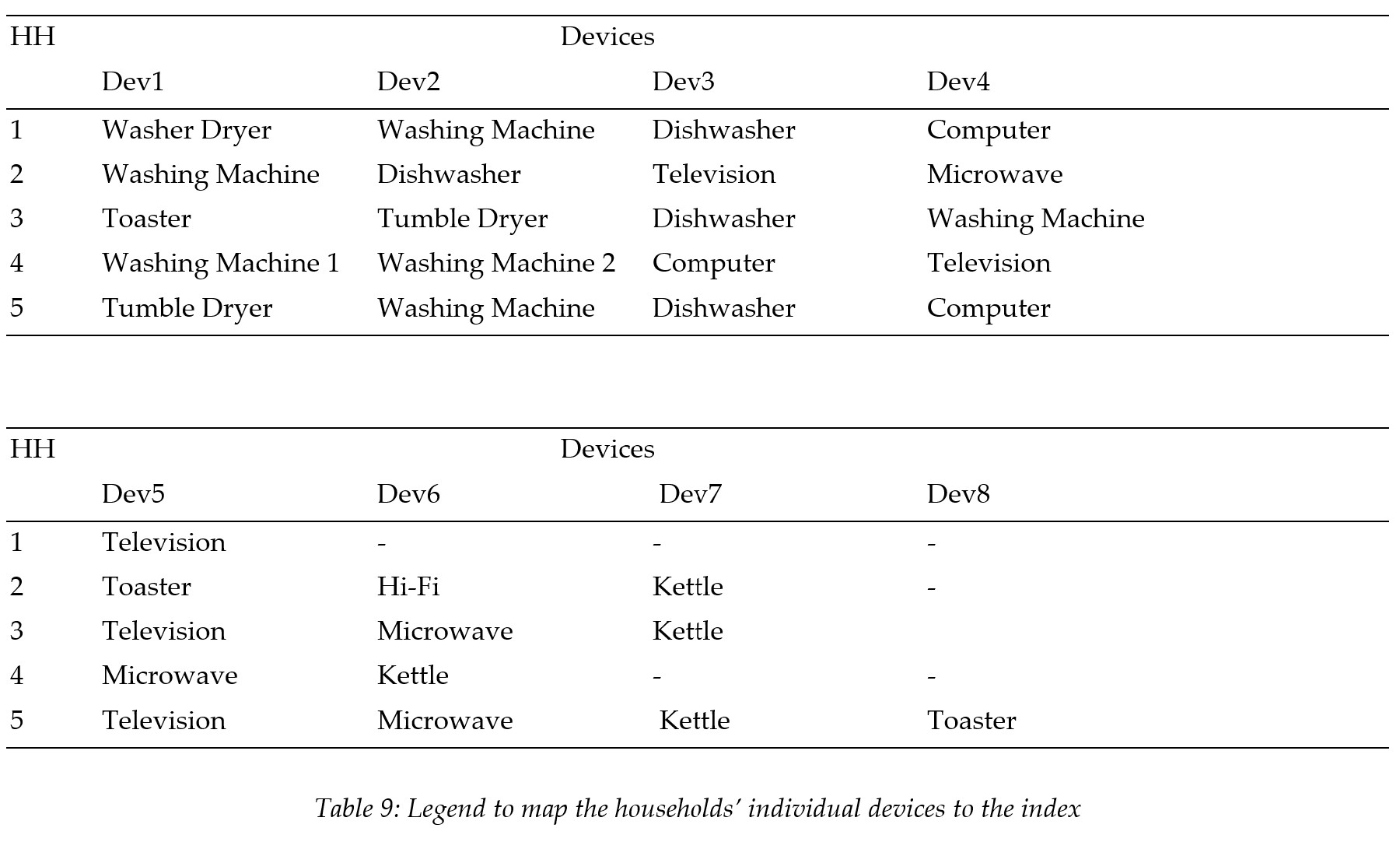}
	\label{}
\end{figure}

\begin{figure}[h!]
	\centering
	\includegraphics[width=0.8\textwidth]{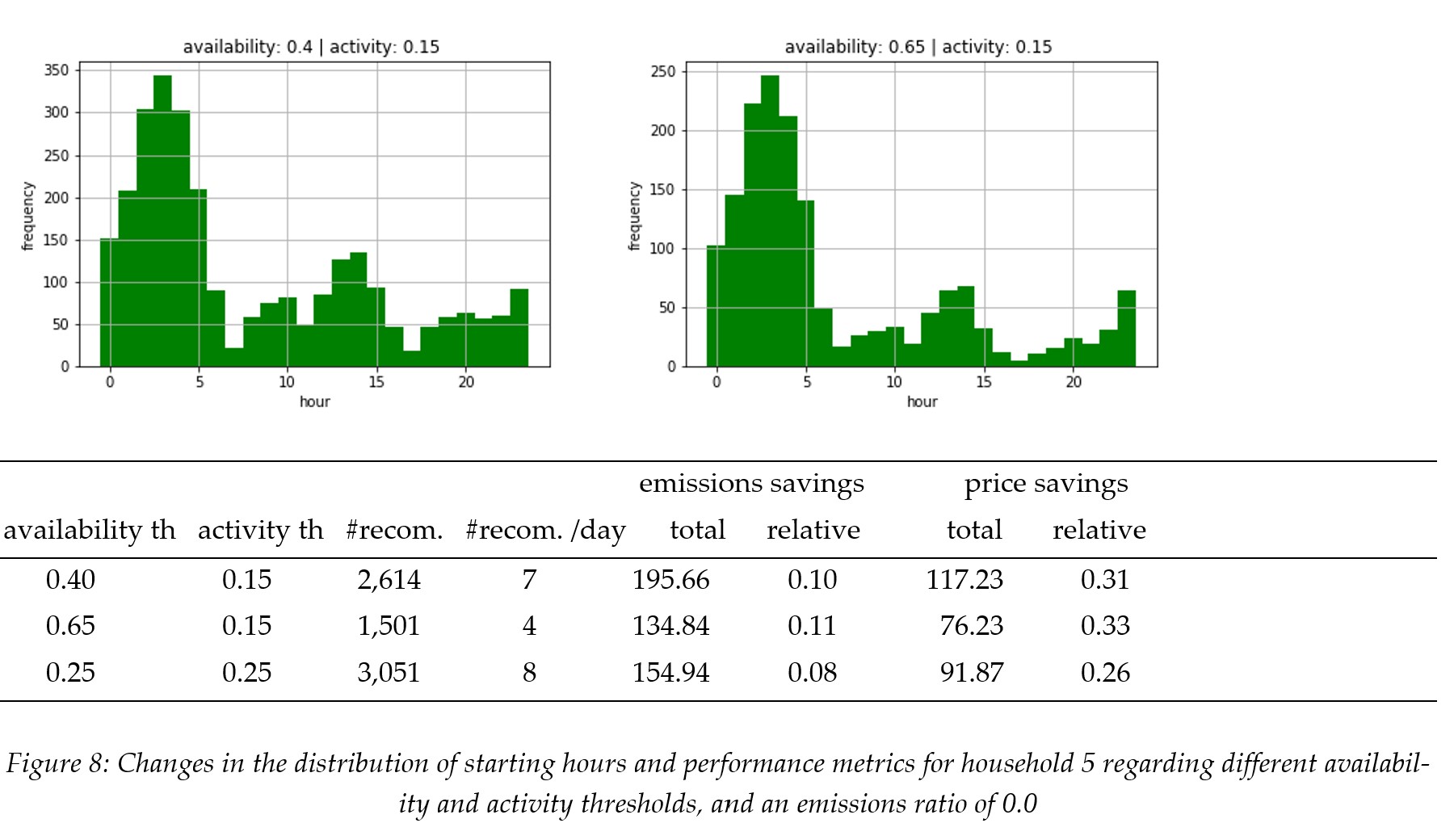}
	\label{}
\end{figure}

\begin{figure}[h!]
	\centering
	\includegraphics[width=0.8\textwidth]{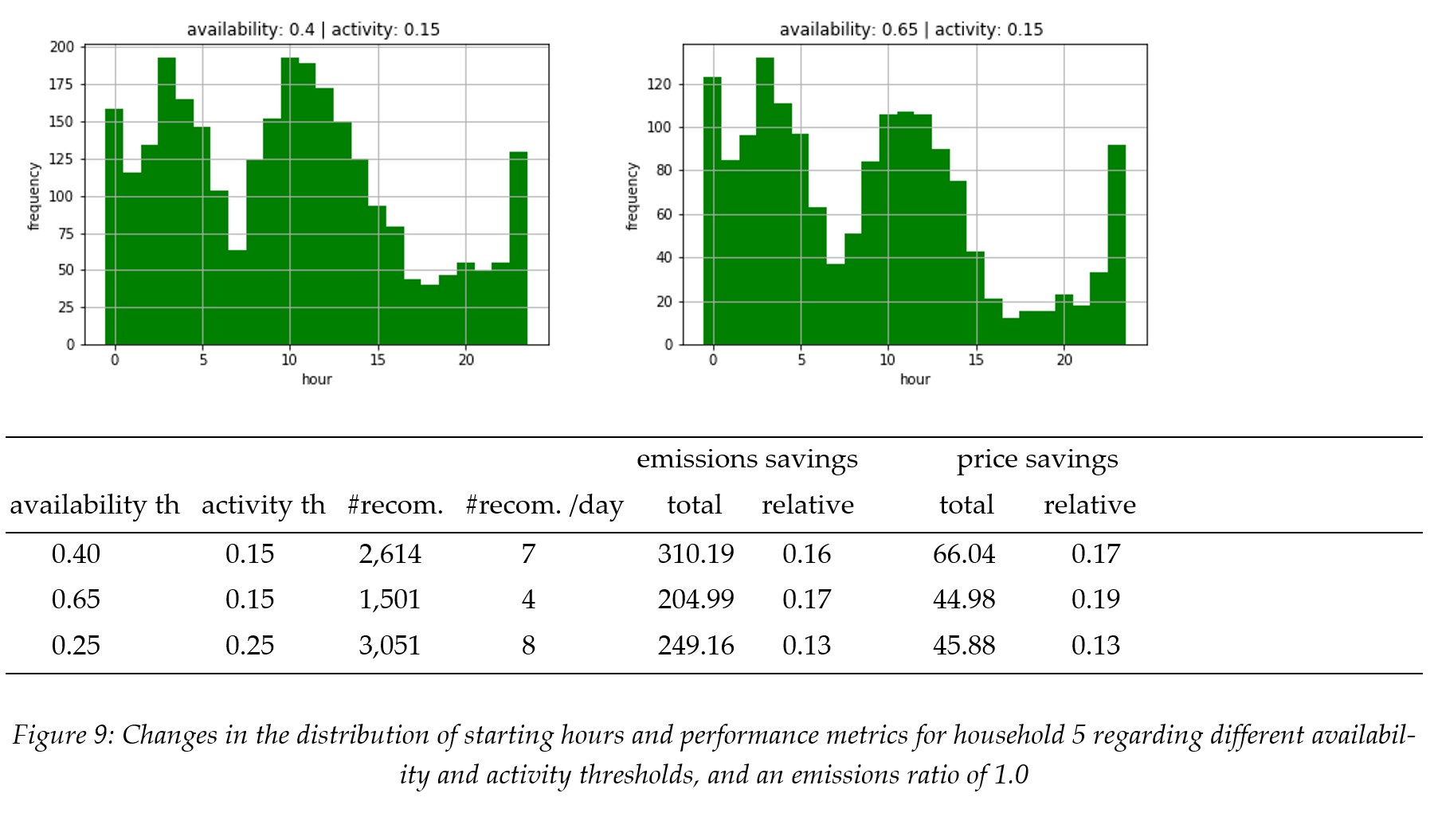}
	\label{}
\end{figure}

\end{document}